\documentclass[prb,twocolumn,showpacs,preprintnumbers,amsmath,amssymb]{revtex4}

\usepackage{graphicx}
\usepackage{dcolumn}
\usepackage{bm}

\newcommand{\cf}[1]{\langle #1 \rangle}            
\newcommand{\no}[1]{: \! #1 \! :}                  
\newcommand{\1}{\openone}                          
\newcommand{\vJ}{\mbox{\boldmath $J$}}             
\newcommand{\vS}{\mbox{\boldmath $S$}}             
\newcommand{\be}{\begin{equation}}
\newcommand{\ee}{\end{equation}}
\newcommand{\bea}{\begin{eqnarray}}
\newcommand{\eea}{\end{eqnarray}}

\newcommand{\vl}{\mbox{\boldmath $\ell$}}
\newcommand{\vn}{\mbox{\boldmath $n$}}
\newcommand{\vR}{\mbox{\boldmath $r$}}
\newcommand{\vA}{\mbox{\boldmath $A$}}

\newcommand{\vLambda}{\mbox{\boldmath $\Lambda$}}
\newcommand{\vsigma}{\mbox{\boldmath $\sigma$}}
\newcommand{\vGamma}{\mbox{\boldmath $\Gamma$}}
\newcommand{\vomega}{\mbox{\boldmath $\omega$}}
\newcommand{\vOmega}{\mbox{\boldmath $\Omega$}}

\newcommand{\vO}{\mbox{\boldmath ${\cal O}$}}
\newcommand{\del}{\partial}
\newcommand{\bml}{\begin{mathletters}}             
\newcommand{\eml}{\end{mathletters} \hspace{-5pt}} 
\newcommand{\Neel}{\mbox{N\'{e}el}}

\begin{document}

\title{Pairing and Density Correlations of Stripe Electrons \\
in a Two-Dimensional Antiferromagnet}

\author{Henrik~Johannesson}

\affiliation{Institute of Theoretical Physics,
Chalmers University of Technology and G\"oteborg University, \\
SE-412 96 G\"oteborg, Sweden}

\author{G.~I.~Japaridze}

\affiliation{Andronikashvili Institute of Physics,
Georgian Academy of Sciences, \\ Tamarashvili str.~6, Tbilisi 380077, Georgia}

\date{January 30, 2003}

\begin{abstract}
We study a one-dimensional electron liquid embedded in a 2D antiferromagnetic insulator,
and coupled to it via a weak antiferromagnetic spin exchange interaction. We argue that this
model may qualitatively capture the physics of a single charge stripe in the cuprates on length-
and time scales shorter than those set by its fluctuation dynamics. Using a local mean-field
approach we identify the low-energy effective theory that describes the
electronic spin sector of the stripe
as that of a sine-Gordon model. We determine its phases
via a perturbative renormalization group analysis.
For realistic values of the model parameters we obtain a
phase characterized by enhanced spin density and composite charge
density wave correlations, coexisting with subleading triplet and
composite singlet pairing correlations. This result is shown to be
independent of the spatial orientation of the stripe on the square
lattice. Slow transverse fluctuations of the stripes tend to
suppress the density correlations, thus promoting the pairing
instabilities. The largest amplitudes for the composite instabilities
appear when the stripe forms an antiphase domain wall in the
antiferromagnet. For twisted spin
alignments the amplitudes decrease and leave room for a new type
of composite pairing correlation, breaking parity but preserving time
reversal symmetry.
\end{abstract}


\maketitle

\newpage


\section{Introduction}
\label{section1}

Extensive experimental studies - including elastic and inelastic neutron scattering 
\cite{Tranquada}, ARPES \cite{Shen}, $\mu$SR \cite{Borsa}, and NMR experiments 
\cite{Suzuki} - have confirmed that {\em stripe formation} is a property common to 
most high-T${_c}$ cuprates. In the underdoped regime, at some critical hole doping, 
the mobile holes segregate into an array of ``stripes'' that slice the copper-oxide 
planes into alternating phase - antiphase antiferromagnetic domains. The stripes 
coexist with superconductivity, but as one enters the overdoped region they begin to 
evaporate, signaling a crossover to a conventional metal with a uniform charge 
distribution. Significantly, stripe phases are observed also in other doped 
antiferromagnets, such as the ``Nickelates'' \cite{NICKELATES} and the ``Manganites'' 
\cite{MANGANITES} (colossal magnetoresistance materials, where the ``stripes'' are 
actually 2D sheets of hole-rich regions). This suggests that stripe formation is a 
robust and generic property of this class of matter. Still, the basic questions why 
stripes form and what role they play for superconductivity in the cuprates remain 
controversial. 

Early mean-field calculations on the 2D Hubbard model \cite{ZaanenGunnarsson} 
suggested that the stripe phase is due to the reduction in kinetic energy of holes
propagating transverse to the stripes. In this approach, however, the possible
connection to superconductivity is left unanswered. In an alternative approach 
\cite{EmeryKivelson}, it is argued that stripes form as a response to the competition 
between long-range Coulomb interactions (which push the holes apart) and short-range
antiferromagnetic interactions (which tend to ``phase separate'' the holes into a 
single region). Within this scenario it has been  argued that a proposed spin gap 
from the undoped domains is transmitted to the stripes via pair hopping (``spin gap 
proximity effect'' \cite{EKZ}), leading to enhanced charge density wave (CDW) as well 
as superconducting pairing correlations along the stripes. For static stripes (as 
seen in e.g. in the Nickelates, or the Nd-doped La$_{2-x}$Sr$_{x}$NdCuO$_4$) the CDW 
correlations dominate. In the presence of transverse stripe fluctuations, however, 
these appear to die out \cite{KFE}, possibly opening a door to superconductivity. 
Other scenarios, where the stripes actually {\em compete} with superconductivity, 
have also been proposed \cite{WhiteScalapino}.

Most of the theoretical attempts to explore the properties of stripes model these as 
a collection of one-dimensional (1D) or quasi-1D electron liquids \cite{Voit}, coupled 
to their neighbors \cite{AHCN}, or to an insulating background, either via pair hopping 
of charge carriers (as in the "spin gap proximity effect" \cite{EKZ}) or by a 
{\em spin exchange}. The various spin-exchange scenarios that have been suggested 
\cite{SachdevShankar,KrotovLeeBalatsky,MGHJ} also predict that a spin gap opens in the spectrum of the 
stripe electrons, signaling enhanced superconducting fluctuations along the stripes. 
In fact, it is common to find a dynamically generated spin gap for a one-dimensional 
electron gas (1DEG) coupled to an active environment 
\cite{EKZ,VarmaEtAl,DagottoEtAl,CastroNeto,EKZb}, of which an antiferromagnet is a 
particular realization 
\cite{SachdevShankar,KrotovLeeBalatsky,MGHJ,ZKE,Fujimoto,Sikkema,Zheng}.

The simplest such model is maybe that of the 1D Kondo-Heisenberg lattice (KHL) which 
consists of a 1DEG interacting weakly with an antiferromagnetic Heisenberg spin-1/2 
chain by a Kondo coupling. Away from half-filling this model has a spin gap 
\cite{Fujimoto,Sikkema} and one thus expects the presence of superconducting 
correlations. Indeed, it was shown recently \cite{ZacharTsvelik,Zachar} 
that the spin gap supports {\em composite} \cite{CompositeOrder1,CGT} odd-frequency odd-parity singlet pairing 
\cite{OddFrequency1} as well as a composite CDW. 
A generalization of this model that may mimic stripe physics more closely is that of 
a 1DEG coupled by a Kondo coupling to {\em two} non-interacting antiferromagnetic 
Heisenberg spin-1/2 chains, together emulating the insulating background in which a stripe 
is embedded. Rather surprisingly, as shown recently \cite{AzariaLecheminant}, 
this generalized model has no spin gap but instead renormalizes to a fixed point 
belonging to the class of chirally stabilized electron liquids \cite{Andrei}. 
Still, the model exhibits the same unconventional pairing instabilities as found for 
the 1D KHL \cite{ZacharTsvelik,Zachar}.

A "strong" interpretation of the results in Refs. 25,26 and 30     
may seem to exclude spin exchange as a 
possible source of the spin gap in the high-T${_c}$ cuprates: odd-frequency pairing 
appears difficult to reconcile with the experimental observation that 
superconductivity in 
these compounds is due to d-wave BCS paired electrons. However,
the recent report \cite{Kaminski} that underdoped BSCCO breaks
time reversal symmetry - in the ''normal'' as well as
the superconducting state - cautions us that the case may not be
closed. The time reversal breaking is seen below a temperature
$T_{gap}$ at which a pseudogap \cite{pseudogapReview} opens,
suggesting that it is connected with some order parameter that
develops enhanced correlations below this characteristic
temperature \cite{OddFrequencyGap}. It has been argued that the pseudogap in the
cuprates may be identified with the amplitude of the pairing
order parameter, with long-range superconducting order appearing
at the onset of global phase coherence (carried by Josephson tunneling
of pairs between the stripes) \cite{Carlson}.  
One may envision a variant of this scenario where spin exchange
between the stripes and their environment (maybe in conjunction
with pair hopping) supports two coexisting types of quasi-one
dimensional pairing correlations below $T_{gap}$, one of which
breaks time reversal. As one approaches the superconducting
transition, the enhanced stripe fluctuations may favor the other
type (which could re-emerge as long-range d-wave order via the
dimensional crossover \cite{Carlson} implied at $T_c$), while
the channel that exhibits time-reversal breaking remains
incoherent, with only finite-range correlations present. Although
speculative only, the viability of this brand of scenario can be
judged only by more closely examining the physics driven by a
stripe-environment spin-exchange interaction. This is the purpose
of our paper.

We shall consider an extended version of the model in Ref. 16,
where a 1D electron liquid (representing a single stripe) is embedded in a
2D antiferromagnetic background, and coupled to it via an
antiferromagnetic spin exchange. We show that this setup leads to
a spin-gap phase for the electrons on the stripe, and we identify its 
leading instabilities. 
We further address the question to what extent the 
instabilities found are sensitive to the relative orientation of the staggered 
magnetizations on each side of the stripe. In the simplest case of a 
{\em site-centered stripe} \cite{TranquadaEtAl} the spin alignment across the stripe 
is antiferromagnetic {\em (phase-antiphase domains)}. However, the alignment is not 
expected to be perfect and it is therefore important to check the stability of the 
spin gap phase w.r.t. deviations from the phase-antiphase orientation of the magnetic 
domains separated by the stripe. In addition, we shall explore the issue whether 
the {\em spatial} orientation of the stripe on the underlying lattice may influence 
the stripe electron dynamics when the dominant interaction with the environment is 
that of a spin exchange. 

This latter question is of particular relevance considering recent experimental 
findings of "diagonal stripes" in the underdoped glassy phase of the cuprates. As 
discovered by Wakimoto {\em et al.} \cite{Wakimoto}, the insulating 
La$_{1.95}$Sr$_{0.05}$CuO$_4$ exhibits sharp two-dimensional elastic magnetic peaks 
at $(\pi \pm \epsilon, \pi \pm \epsilon)$ (in tetragonal square lattice notation, with
$\epsilon \sim x \approx 0.05, x$ being the doping level). Assuming that the magnetic
peaks are associated with charge stripe order \cite{TranquadaReview}, this implies that
static stripes run along the diagonal of the square  Cu$^{2+}$ lattices that make up 
the CuO-planes in this compound. This is in exact  analogy to the diagonal static 
stripe structure seen (and theoretically predicted   \cite{ZaanenGunnarsson}) in the 
insulating Nickelate  La$_{2-x}$Sr$_{x}$NiO$_{4+x}$, but {\em different} from the 
structure in {\em superconducting} La$_{2-x}$Sr$_{0.05}$CuO$_4$ (with $x > 0.05$) 
where the stripes are oriented along the copper-oxide bonds (``collinear stripes''). 
Very recently, these finding were extended to the full insulating spin glass phase in 
La$_{2-x}$Sr$_{x}$CuO$_{4}$  ($0.02 \le x \le 0.055$) \cite{Matsuda}. Thus, the onset 
of  superconductivity in the low-temperature underdoped region appears to proceed via a {\em stripe 
rotation} by $45^{\circ}$, from a {\em diagonal} to a  {\em collinear} stripe 
configuration. In the case of a collinear (site-centered) structure, a 
stripe is embedded in a local  {\em antiferromagnetic} background. In 
contrast, in the diagonal structure the stripe electrons experience a 
local {\em ferromagnetic} environment. In both cases the  
antiferromagnetic domains are $\pi$-shifted {\em across} the stripes, and it is 
{\em a priori} not clear whether the different local orderings 
{\em along} the 
stripes influence the electron dynamics differently. 
To clarify the situation 
requires a careful study, and we here make a first attempt on it. \\

To isolate the core of the problem we shall make a few simplifying assumptions: \\  
{\bf (i)} \ We study the electron dynamics on a {\em single} stripe, and, in 
the first part of our analysis, neglect its 
possible coupling to neighboring stripes. Moreover,  
the stripe is taken to be {\em static}. This implies that for a 
fluctuating stripe (as typically  seen in a superconducting phase) we can only hope 
to capture processes on length- and time scales shorter than those set by 
its 
fluctuation dynamics. This is expected to be much slower than the 
dynamics of charge 
carriers along the stripe: The latter appear on an energy scale
$\sim 1$eV, whereas the stripe fluctuations are coordinated with
those of the localized spins, at a scale $\sim$
1- 10meV. Having obtained the characteristic features of a single static stripe
we then add ''by hand'' the transverse fluctuations and interstripe couplings,
and study their effect on the pairing- and density correlations of the stripe
electrons. 
\\
{\bf (ii)} \ We model the stripe - at Fermi momenta incommensurate with the 
underlying lattice - as a {\em one-dimensional metallic wire}. Thus, we assume that 
the disorder (from e.g. dopant potentials) is sufficiently weak so that localization 
effects set in on length scales  much larger than those that we probe here. \\ 
{\bf (iii)} \  There is an important distinction between site-centered and 
bond-centered stripes \cite{TranquadaEtAl}: The spin alignment across an anti-phase 
domain wall is antiferromagnetic for a site-centered stripe and ferromagnetic for a 
bond-centered stripe (which has a finite width). Motivated by recent theoretical 
work \cite{Zacher} where bond-centered configurations appear to be inconsistent with 
ARPES studies of La$_{2-x}$Sr$_{x}$CuO$_{4}$ \cite{Ino}, we here focus on the 
simplest case of a site-centered charge stripe, to be described by a 1D Hubbard 
model. We shall explore elsewhere the case of bond-centered stripes, building on the 
corresponding analysis by Krotov, Lee and Balatsky \cite{KrotovLeeBalatsky} of a 
Hubbard ladder in an antiferromagnetic environment. \\ 
{\bf (iv)} \ As suggested by neutron scattering data on the relevant materials 
\cite{Tranquada}, the environment is {\em N\'{e}el ordered} up to some characteristic 
scale (which, in the relevant temperature range, is much larger than the linear 
dimension of a stripe), with a $\pi$-shift 
across the stripe when this is site centered (phase - antiphase domains). 
In our formal analysis we depict each N\'{e}el ordered domain as a semi-infinite 2D 
Heisenberg antiferromagnet, and ignore possible topological effects that may be 
present for finite-width insulating domains, or "spin ladders" \cite{SierraReview}. 
We shall give precise estimates for the range of validity of this approximation, 
thus establishing its physical relevance. \\ 
{\bf (v)} \  As we have already discussed, we couple the stripe electrons to its 
insulating environment exclusively through a {\em spin exchange interaction}. Given that the 
Fermi momentum of the stripe is incommensurate with that of any low-lying excitation 
of the environment, excursions of single charge carriers is a process that violates 
momentum conservation, and hence is suppressed 
(on the time scales defined in (i)). Pair hopping is still allowed, 
provided that the pair carries zero total momentum. As suggested by the analysis in Ref. 9,
pair hopping is favored as a dominant process when the low-lying spin 
excitations of the environment are gapped \cite{chargegap}. When such a gap is absent, as is the case 
when the environment is N\'{e}el ordered, the virtual hybridization between 
delocalized levels on the stripe and the localized levels in the environment produces 
an effective spin exchange that is expected to compete effectively with pair hopping. 
Here, we focus on the effect of the spin exchange. \\  
{\bf (vi)} \ We limit our attention to the physically relevant case of a 
{\em weak spin exchange} $J_K$ between stripe and environment, 
$0 < J_K \ll J_H$,
where $J_H$ is the antiferromagnetic exchange 
between the localized spins {\em in} the environment. This allows us to treat the
problem in a continuum limit \cite{Sikkema}. \\
{\bf (vii)} \ Finally, we stress that finite-size or boundary effects  
of the 1D electrons \cite{Fabrizio} are {\em not} included in our analysis. As the stripes in the 
cuprates are mesoscopic structures \cite{STRIPEDYNAMICS}, these effects {\em should}
in principle be taken into account. However, as they are not expected to qualitatively change
the conclusions arrived at in the large-distance limit considered here, we leave this
study for the future.

Clearly, by assumptions (i) - (vii) we loose several facets of the full problem. 
Still, we believe that our "stripped-down" approach has its merits: Not only does 
it  isolate and expose a crucial element of "stripe physics", but as we shall show, it 
allows us to perform a {\em well-controlled} analytical study, producing results that 
can be taken as a reliable starting point for more realistic studies. Moreover, the 
problem as defined by (i)-(vii) is important in its own right, and is of relevance to 
the more general issue of one-dimensional electron liquids in active environments 
\cite{EKZ,VarmaEtAl,DagottoEtAl,CastroNeto,EKZb}. This is a central problem in the 
theory of correlated electrons, motivated by experiments on quasi-1D organic 
conductors \cite{BECHGAARDSALTS}, nanowires  \cite{Auslaender}, 
and edge states in quantum Hall systems \cite{Grayson}. 

Given the assumptions (i) - (vii), we model the stripe by an
extended $(U-V)$ Hubbard chain, weakly 
coupled to a phase - antiphase antiferromagnetic environment by a Kondo lattice 
interaction. Treating the Hubbard chain via standard bosonization while 
describing the 
environment by a nonlinear $\sigma$-model, we follow the approach introduced in
Ref. 16 and exploit the symmetry breaking in the magnetic
environment to ''absorb'' the Kondo lattice interaction as an effective spin-spin 
interaction among the stripe electrons. In this way we obtain an effective low-energy 
model for the stripe electrons - decoupled from the environment -  and accessible to a 
well-controlled perturbative renormalization group analysis. This allows us to pinpoint 
the dynamic instabilities in the low-energy, weak-coupling $(J_K \ll J_H < 
\left| U \right|, \left| V \right|)$ limit. 

Our most important results can be summarized: 
\begin{itemize}
\item{For realistic values of the model parameters, and with a
phase-antiphase $\Neel$-configuration across the stripe,
an electronic spin gap
opens on the stripe with a spin-density and a composite charge-density
wave as the leading instabilities. The subleading instability is
that of conventional triplet pairing, coexisting with 
composite singlet pairing (which breaks parity and time
reversal). Slow transverse stripe fluctuations tend to suppress the
density correlations, thus promoting the pairing instabilities.}
\item{The low-energy physics is insensitive to the
spatial orientation of the stripe on the lattice: The results
summarized above hold for both collinear and diagonal stripes (with the
possible exception that the composite singlet pairing is suppressed for
a diagonal stripe).}
\item{The instabilities found for the phase-antiphase
$\Neel$-configuration are still present when the relative
orientation of the staggered magnetizations on the respective
sides of a (collinear) stripe 
has been twisted by an arbitrary angle. 
In addition, the twist allows for a novel type of composite pairing correlations
to appear, respecting
time reversal but breaking parity.}
\end{itemize}

The paper is organized as follows: In Sec. II we introduce the lattice models for 
site-centered collinear and diagonal stripes coupled to a phase-antiphase environment 
by a Kondo interaction, and derive the corresponding low-energy effective actions. In 
Sec. III we perform an RG analysis and identify the order parameter correlations along 
the stripes that get enhanced by the spin exchange. In Sec IV we then study - for the 
case of a collinear stripe - the stability of these correlations w.r.t. perturbations 
of  the relative orientation of the spin alignments across the stripe. Sec. IV, finally, 
contains a summary and a brief discussion of our results. 
Throughout the paper we try 
to supply sufficient information to make the analysis accessible also to the non-expert.

\section{The Model}

For clarity, we shall treat the collinear and diagonal stripe configurations separately. 
The effective low-energy theories that emerge in the two cases are 
essentially the same (with 
certain provisos), but to arrive at this result requires some care. 
Much of the
analysis builds upon well-known results, but in order to make the exposition self-contained
we outline the most important points. Also, some key elements are new or need
particular attention. 

\subsection{Collinear Stripes}

We represent the stripe (running, say, along the x-direction of a square lattice) by an 
extended ($U-V$) Hubbard chain, $H_{Hubbard}$, coupled via a Kondo lattice interaction 
$H_{Kondo}$ to the {\em nearest} localized
spins on each side of the stripe. These spins, like the rest of the 
localized spins, interact mutually via an antiferromagnetic
nearest-neighbor Heisenberg spin exchange $H_{AFM}$, and
reside in one of the two semi-infinite antiferromagnetic domains that surround the 
stripe, denoted $A$ and $B$ respectively (see FIG. 1). 

\begin{figure}[tbh]
\begin{center}
\includegraphics[width=0.4\textwidth]{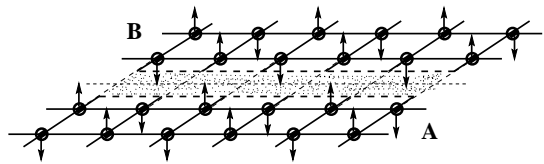}
\caption[COLLINEAR]{Collinear stripe structure.}
\label{FIG1}
\end{center}
\end{figure}

The $A$ and $B$ domains are assumed
to be antiferromagnetically ordered, and correlated via a $\pi$-shift across the stripe 
{\em (phase - antiphase domains)} but there is no direct interaction between $A$ and $B$
spins  \cite{FOOTNOTE1}. Thus, we study the lattice model
\begin{equation}
H = H_{Hubbard} + \sum_{i=A,B} (H_{AFM}^{(i)} + H_{Kondo}^{(i)}),
\label{MODEL}
\end{equation}
where
\begin{eqnarray}
H_{Hubbard} & = & -t \sum_{r, \, \alpha}(c_{r+1,\alpha}^{\dagger}c_{r,\alpha} 
+ h.c.) \nonumber \\ 
& + & U \sum_{r}\hat{n}_{r,\uparrow}\hat{n}_{r,\downarrow}  
+ V\sum_{r} \hat{n}_{r}\hat{n}_{r+1} \, ,\label{HUBBARD}
\end{eqnarray}
\begin{multline}
H_{AFM}^{(i)} = J_H \sum_{r, \, j^{(i)}} ( \vS_{r,j^{(i)}}\cdot
\vS_{r+1,j^{(i)}} + 
 \vS_{r,j^{(i)}} \cdot \vS_{r,j^{(i)} +1} ), \\			 
 J_H > 0,  \ \ i = A, B,  \label{HEISENBERG}
\end{multline}
\begin{multline}
 \label{KONDOLATTICE}
H_{Kondo} = J_K \sum_{r, \,\alpha,\beta} c_{r,\alpha}^{\dagger}\vsigma_{\alpha \beta}
c_{r, \beta} 
\cdot ( \vS_{r,\,j^{(A)} = 1} + \vS_{r,\,j^{(B)} = 1}),  \\ 
 J_K > 0.
\end{multline}
Here $c_{r,\alpha}$ is a stripe electron operator at site $r$ with spin index $\alpha 
= \uparrow, \downarrow$, $\hat{n}_{r,\alpha} = c^{\dagger}_{r,\alpha} c_{r,\alpha}$ is the 
number operator, $\hat{n}_{r}=\sum_{\alpha}\hat{n}_{r,\alpha}$ is the density operator, and 
$\vS_{r, j^{(i)}}$ is the operator for a localized spin at a lattice site with coordinate 
$r \ (j^{(i)} \ge 1)$ in a direction parallel with (transverse to) the stripe. 
The vector of Pauli matrices is denoted by $\vsigma$, and we have absorbed a factor of
1/2 into the coupling constant $J_K$.
Note that we have included a nearest-neighbor interaction in
$H_{Hubbard}$ to qualitatively account for the poor screening of
the Coulomb interaction from the insulating environment. 
The {\em Coulomb-driven} on-site and nearest-neighbor coupling 
constants are typically repulsive $U,V>0$. However, in what
follows we will treat these parameters as 
effective (phenomenological) ones, and assume that they include all possible contributions 
and renormalizations coming from the interaction between stripe electrons 
and the {\em nonmagnetic} 
degrees of freedom of the environment, such as the electron-phonon 
coupling, or coupling to 
other electronic subsystems in the environment 
\cite{MRR}. As implicit 
in Eq.~(\ref{KONDOLATTICE}), we use the convention that the transverse coordinates take values 
$j^{(A)} = j^{(B)} = 1$ on the $A$- and $B$-arrays adjacent to the stripe. When convenient, we use 
the compact notation $\vS^{(i)}_r \equiv \vS_{r,  \, j^{(i)}}
(i=A, B)$ for the spins on these arrays. 
By assumption (vi) in Sec. I, the antiferromagnetic Kondo lattice coupling 
$J_K$ is weak, i.e. $J_K \ll J_H, t$, allowing us to keep only the low-energy sectors 
of the stripe and the antiferromagnetic domains when analyzing its effect. 

The model in Eq.~(\ref{MODEL}) is a modified version of that in Ref. 16
by having a coupling of {\em two} semi-infinite 2D antiferromagnetic domains to the stripe, 
one on each side of it, in this respect mimicking
the geometry ''seen'' by a real stripe. As a consequence, with the 
assumption that the N\'{e}el order 
of the $A$ and $B$ domains are $\pi-$phase shifted relative to each other, we will be 
able to treat the metallic as well as the Mott insulating case (half-filled Hubbard 
band) within the same formalism. This is different from the model in Ref. 16, where the 
assumption of a  metallic stripe was crucial.  Note that by turning off $U$ and $V$ in 
(\ref{HUBBARD}) and keeping only one array of localized spins, say, in the $A$-domain, 
the Hamiltonian in (\ref{MODEL}) collapses to the one-dimensional {\em
Heisenberg-Kondo-lattice model} (HKL) \cite{TsvelikReview}. This model has recently 
attracted a great deal of attention \cite{Fujimoto,Sikkema} and we shall connect back 
to it when discussing our results in Sec. IV.  We should here emphasize that having 
"built in" the presence of stripes into the model, we cannot address the issue of what 
actually triggers the stripe formation. For this, one must turn to other approaches, 
some of them mentioned above \cite{ZaanenGunnarsson,EmeryKivelson,WhiteScalapino}. A 
recent attack on the problem of stripe formation has made use of a {\em spin-fermion 
model} \, \cite{Moreo} which has the same Hamiltonian structure as (\ref{MODEL}),
but with the difference that there is no constraint on the mobility of the 
electrons in (\ref{HUBBARD}). In other words, the doped holes are now free to hop 
around on the 2D lattice. Monte Carlo simulations on the model suggest that the holes 
self-organize into one-dimensional charge stripes separating insulating spin domains 
(phase shifted by $\pi$ across the stripes). This is the starting point when writing
down our model in (\ref{MODEL}).

Given the Hamiltonian in (\ref{MODEL}), its partition function can be written as a path 
integral,
\begin{equation}
Z=\int {\cal D}[c]{\cal 
D}[c^{\dagger}] {\cal D}[{\bf \Omega}_A]{\cal D}[{\bf \Omega}_B] \, e^{-{\cal S}[ c^{\dagger}, 
\,c, \,{\bf \scriptstyle \Omega}_A,\, {\bf \scriptstyle \Omega}_B ]}\,, 
\label{PARTITIONFUNCTION}
\end{equation}
with a Euclidean action ${\cal S}$. The electron operators are here simulated by
Grassmann numbers $(c^\dagger_{r, \alpha}, c_{r, \alpha})$, while the role of the
localized spin operators $\vS_{r, j^{(i)}}$ are played by vectors 
$S {\bf \Omega}_{r, j^{(i)}} \,
(i \!= \!A, B)$ which parameterize states in a coherent spin representation \cite{Fradkin}.
Note that we have here used the short-hand notation $c \equiv \{c_{r, \alpha}\}$  and 
${\bf \Omega}_i \equiv \{ {\bf \Omega}_{r,j^{(i)}}\} , i \!= \!A, B$. 
In the limit of large spin, $S \rightarrow \infty$, only diagonal matrix 
elements of a spin Hamiltonian survive in this representation. This makes 
the large-$S$ coherent state representation an efficient tool to mold a quantum partition 
function into a path integral. To retain quantum effects, however, present for physical 
values of the spin ($S=1/2$ in the case of the cuprates), it is crucial to keep also non-diagonal 
matrix elements in the construction of ${\cal }S$. These produce a sum over Berry 
phases and contain a memory of the intrinsic quantum nature of the spins. The procedure is 
standard \cite{TEHJ}, and one obtains the action 
\begin{eqnarray}
{\cal S} & = &  
\int_0^{\beta} d\tau \, 
\left( \sum_{r} c^\dagger_{r,\alpha} \partial_{\tau} c_{r,\alpha} + 
H(c^\dagger,c,S{\mbox{\boldmath $\Omega$}}_A, S{\mbox{\boldmath $\Omega$}}_B)\right) \nonumber \\
& + &  iS \sum_{i=A,B} \ \sum_{r,j^{(i)}} \Phi_{r, j^{(i)}}, 
\label{ACTION}
\end{eqnarray}
where $\tau$ corresponds to inverse 
temperature so that $ 0 < \tau < \beta$ and the spin (Grassmann) fields are
periodic (antiperiodic) in $\tau$. For the purpose of studying the low-energy dynamics 
we confine our attention to the zero-temperature limit $(\beta \rightarrow \infty$). The third
term in (\ref{ACTION}) is precisely the sum over Berry phases  
\begin{equation}
\Phi_{\vR_i} =
\oint_{{\Gamma}_{\vR_i}} d{\bf \Omega}_{{\vR}_i} \cdot 
{\bf A}({\bf \Omega}_{{\vR}_i}), \label{BERRYPHASE}
\end{equation}
one for each spin attached at site $(r, j^{(i)}) \equiv {\vR}_i , i=A,B$. Here ${\Gamma}_{{\vR}_i}$ is 
the closed loop traced out by ${\bf \Omega}_{{\vR}_i}$ in the interval 
$[0, \beta \rightarrow \infty]$, with  ${\bf A}({\bf\Omega}_{{\vR}_i})=(1\!-\! \mbox{cos}\,
\theta_i)(\mbox{sin}\,\theta_i)^{-1}\varphi_i$ at each site 
$\vR_i$ playing the role of a vector potential of a unit magnetic monopole located at 
the center of the sphere $\left| {\bf \Omega}_{\vR_i} \right| = 1$, parameterized by the spherical 
angles $\theta_i$ and $\phi_i$.  The "instantaneous" Hamiltonian term 
$H(c^\dagger, c, S{\mbox{\boldmath $\Omega$}}_A, S{\mbox{\boldmath $\Omega$}}_B)$ in 
(\ref{ACTION}) acts at (imaginary) time  slice $\tau$ and is obtained from (\ref{MODEL}) by 
replacing electron and spin operators by the corresponding Grassmann fields 
$(c^{\dagger}_{r,\alpha}, c_{r,\alpha})$ and classical vectors 
$(S {\bf \Omega}_{{\vR}_A} , S {\bf \Omega}_{{\vR}_B})$, respectively.

Next, to obtain the low-energy continuum version of the Hamiltonian in (\ref{MODEL})
we shall first review the standard constructions for a Hubbard chain and a 2D Heisenberg
model, and then elaborate on the more intricate Kondo-lattice interaction
which couples the two subsystems.

\subsubsection{1D Electron Chain: Low-Energy Theory}

The low-energy (field theory) approach to the 1D (extended) Hubbard model in (\ref{HUBBARD}) 
is based on the assumption of {\em weak} electron-electron interactions. 
Thus, assuming that
$\left| U \right|, \left| V \right| \, \ll t$, we linearize the spectrum around the two Fermi points 
$\pm k_{F}$ ($k_{F} =n_e\pi/2a_0$, where $a_0$ is the lattice spacing and $n_e$ is the electron 
density), and decompose the original lattice operators into right-moving
$(R_{\alpha})$ and left-moving $(L_{\alpha})$ chiral 
components:
\begin{eqnarray} \label{LINEARIZATION}
c_{r,\alpha} & \rightarrow & \mbox{e}^{ik_Fra_0} R_{\alpha}(r) + \mbox{e}^{-ik_Fra_0}
L_{\alpha}(r) \nonumber \\
& \rightarrow & \sqrt{a_{0}}\left[\mbox{e}^{ik_Fx}R_{\alpha}(x) 
+ \mbox{e}^{-ik_Fx}L_{\alpha}(x)\right],  
\end{eqnarray}
where in the second line we have taken the continuum limit $ra_0 \rightarrow x$.
Defining local charge and spin densities
\bea
J_{R} &=& \no{R^{\dagger}_{\alpha}R_{\alpha}}, \qquad J_{L}= \no{L^{\dagger}_{\alpha}L_{\alpha}}, \, \label{chargedensity} \\
\vJ_{R} &=& \, \no{\frac{1}{2}R^{\dagger}_{\alpha} \vsigma_{\alpha \beta} R_{\beta}}, \qquad 
\vJ_{L} = \, \no{\frac{1}{2}L^{\dagger}_{\alpha} \vsigma_{\alpha \beta} L_{\beta}},  \label{spindensity}  \
\eea
with repeated spin indices summed over, and with the normal ordering $\no{...}$ taken w.r.t. the ground state of the free system, it 
is now straightforward to write down 
the low-energy continuum version of the $U-V$ Hubbard chain 
in (\ref{HUBBARD}). The weak interaction preserves the important property of spin-charge 
separation, and one can write the theory on the form $H_{Hubbard} = H_{c}+H_{s}$, where \cite{Emery}: 
\begin{widetext}
\begin{equation} \label{HUBBARDCONTINUUMcharge}
H_{c}=\frac{\pi v_c}{2} \int \, \mbox{d}x \left\{ \no{J_{R}J_{R}} + \no{J_{L}J_{L}} 
-\, g_{0c} J_{L}J_{R} 
- g_{0u}\delta_{{\small 1} n_e} 
(R^{\dagger}_{\uparrow}R^{\dagger}_{\downarrow}L_{\uparrow}L_{\downarrow} + h.c.)\right\},
\end{equation}
\begin{equation} \label{HUBBARDCONTINUUMspin}
H_{s}= 2\pi \tilde{v}_{s}\int \, \mbox{d}x \Big\{ \no{J^{z}_{R}J^{z}_{R}} + \no{J^{z}_{L}J^{z}_{L}} -\, \tilde{g}_{0s} J^{z}_{L}J^{z}_{R} - 
\tilde{g}_{0\perp}(J_{L}^{x}J_{R}^{x} + J_{L}^{y}J_{R}^{y})\Big\}.
\end{equation}
\end{widetext}
The velocities of the charge $(c)$ and spin $(s)$
excitations, governed by $H_c$ and $H_s$ respectively, are given by 
\begin{equation}  \label{velocities}
v_c = v_F + \frac{a_0(U+6V)}{2\pi} \, , \ \ \ \ \ 
\tilde{v}_s = v_F - \frac{a_0(U-2V)}{2\pi}\, ,
\end{equation}
with $v_{F}= 2 a_{0}t\mbox{sin}(\pi n_e/2)$ the Fermi velocity. 
The small dimensionless coupling constants in 
(\ref{HUBBARDCONTINUUMcharge}) and
(\ref{HUBBARDCONTINUUMspin}) are given by 
\begin{eqnarray}
g_{0c}&=&-a_0(U+6V)/\pi v_{F}, \nonumber \\
g_{0u} &=& -a_0(U-2V)/\pi v_{F}, 
\label{COUPLINGCONSTANTS}\\
\tilde{g}_{0s}&=&\tilde{g}_{0\perp}=a_0(U-2V)/\pi {v}_{F}. \nonumber
\end{eqnarray}
The Kronecker delta multiplying the Umklapp term in (\ref{HUBBARDCONTINUUMcharge}) signifies that this term survives the phase 
fluctuations (originating from the chiral decomposition in ({\ref{LINEARIZATION})) only for 
a half-filled electron band $(n_e = 1)$ \cite{Emery}. The transverse component of the spin 
current coupling in (\ref{HUBBARDCONTINUUMspin}),
\begin{equation}
\tilde{g}_{0\perp}(J_{L}^{x} J_{R}^{x} + J_{L}^{y} J_{R}^{y})= 
-\frac{1}{2}\tilde{g}_{0\perp}
(R^{\dagger}_{\uparrow}L_{\uparrow} L^{\dagger}_{\downarrow}R_{\downarrow} + h.c.),  
\label{BACKCONVERSION}
\end{equation}
describe backscattering of electrons. We have marked the
spin-sector parameters in (\ref{HUBBARDCONTINUUMspin}), (\ref{velocities}) and 
(\ref{COUPLINGCONSTANTS}) by a ''tilde'' as a reminder
that these will be modified when coupling the stripe to the
localized spins in the environment. How this comes about is
discussed next.

\subsubsection{Phase - Antiphase Antiferromagnetic Domains: Low-Energy Theory}

In the presence of antiferromagnetic correlations in the insulating domains - as seen 
experimentally in the stripe materials \cite{Tranquada} - the partition function 
in (\ref{PARTITIONFUNCTION}) at low energies is dominated by paths with 
\begin{equation}
{\mbox{\boldmath $\Omega$}}_{{\small \vR}_i} = 
\gamma_{{\small \vR}_i} (-1)^{\delta_{iB}}\sqrt{1- a_0^2 \vl^2({\small \vR}_i})\,  
\vn({\small \vR}_i) 
+ a_0 \vl({\small \vR}_i). \label{HALDANEANSATZ}
\end{equation}
Here $\gamma_{{\small \vR}_i} = \pm 1$ is the parity of the sublattice to which the site 
${\small \vR}_i$ belongs, the unit vector $\vn$ (suppressing the coordinate ${\small \vR}_i$ for 
ease of notation) represents the local direction of the N\'{e}el order parameter field, and 
$a_0 \vl$ is a small orthogonal ferromagnetic fluctuation component, i.e. 
$\left| a_0\vl \right| \ll 1$, with $\vn \cdot \vl = 0$. The phase factor 
$(-1)^{\delta_{i B}}$ in (\ref{HALDANEANSATZ}) appears because from now on we take the 
N\'{e}el fields in the $A$ and $B$ domains to be $\pi$-shifted relative 
to each other. (The choice of reference vector in the staggering factor can be made arbitrarily, with no
effect on the physics.) We here note that for a site-centered stripe
embedded in a
spin-1/2 environment current estimates predict that this is a
viable assumption for
stripe electron densities $n_e < 0.6$ in the limit where $J_K \sim J_H$ \cite{Zachar2,LiuFradkin}.
However, for $J_K \ll J_H$, as assumed here, the critical density is expected to be larger.

With $\vn$ a slowly varying smooth field, (\ref{HALDANEANSATZ}) spells 
out the assumption of finite-range antiferromagnetic order. We should stress that 
$\vn$ and $\vl$ are 
taken to be independent fields, constrained only by the orthogonality condition. This 
implies a doubling of degrees of freedom, which in principle should be corrected for 
when regularizing the theory. However, for our present purpose, to pinpoint the leading 
instabilities of the stripe electron dynamics due to the interaction with the 
antiferromagnetic domains, this issue is immaterial. 
The normalization of ${\mbox{\boldmath $\Omega$}}$ in (\ref{HALDANEANSATZ}) is 
only preserved up to ${\cal O}(a_0^2)$, but this is sufficient since we are interested in 
the long-wave length limit. In this limit we let the lattice spacing $a_0$ in the $x-$direction 
(parallel to the stripe) go to zero, expand all terms in the action $S$ which contain the 
spin fields up to ${\cal O}(a_0^2)$, and then do the replacements $a_0 \sum_{r} \rightarrow$  
$\int dx$, $\vn_{{\small \vR}_i}(\tau) \rightarrow \vn_{j^{(i)}}(\tau, x),$ and 
$\vl_{{\small \vR}_i}(\tau) 
\rightarrow \vl_{j^{(i)}}(\tau, x)$. The result is a field theory for the independent orthogonal 
fields $\vn$ and $\vl$, with an action  
\begin{widetext}
\begin{eqnarray}
{\cal S}_i [{\scriptstyle \vn}, {\scriptstyle \vl} ]  & = &\frac{S^2 J_H}{2} \sum_{j^{(i)}} 
\int dx \int_0^{\infty} d\tau \Big[ \frac{1}{a_0} \Big(\vn_{j^{(i)} +1}(\tau, x) - 
\vn_{j^{(i)}}(\tau, x)\Big)^2 + 8a_0\vl^2_{j^{(i)}}(\tau, x) \Big]   \nonumber \\
& - & iS \sum_{j^{(i)}} \int dx \int_0^{\infty } d\tau 
\Big(\vn_{j^{(i)}}(\tau, x) \times \partial_{\tau} \vn_{j^{(i)}}(\tau, x)\Big) 
\cdot \vl_{j^{(i)}}(\tau, x)  + {\cal S}_{i, Berry} [{\scriptstyle \vn}], \ \ i = A, B, \label{SPINACTION} 
\end{eqnarray}
\end{widetext}
with  
\begin{equation} \label{BERRYSUM}
{\cal S}_{i, Berry} [{\scriptstyle \vn}]
\! = \! iS \sum_{{\small \vR}_i} \gamma_{{\small \vR}_i} (-1)^{\delta_{i B}} \int _0^{\infty} 
d\tau \vA(\tau, {\small \vR}_i) \cdot \partial_{\tau} \vn(\tau, {\small \vR}_i). 
\end{equation}
In the standard approach to the 2D antiferromagnet \cite{Fradkin} one would now integrate 
out the rapidly fluctuating $\vl$-field from (\ref{SPINACTION}). After taking a continuum 
limit in the $y-$direction this would produce the familiar nonlinear $\sigma$ model (NL$\sigma$M), 
with an added sum over Berry phases, describing the slow long-wavelength dynamics of the N\'{e}el 
order-parameter field $\vn$. In the present case, however, the localized spins adjacent to the 
stripe enter also in the Kondo lattice interaction (\ref{KONDOLATTICE}), and this must be taken 
into account before one attempts to integrate out the $\vl-$field. This problem is addressed 
next.

\subsubsection{Kondo Lattice Interaction: Mean-Field Decoupling} 

Using (\ref{LINEARIZATION}) to write the electron spin density in (\ref{KONDOLATTICE}) in terms 
of the chiral fields, and then replacing the spin operators $\vS_{r, j^{(i)}}$ in 
(\ref{KONDOLATTICE}) by the corresponding vectors $S {\vOmega}_{r, j^{(i)}}$, decomposing 
these as in (\ref{HALDANEANSATZ}), we obtain, expanding to ${\cal O}(a^2_0)$,
\begin{equation}
H_{Kondo} = H_{{\small \vl}} + H_{{\small \vn}},  \label{KONDODECOMP}
\end{equation}
where
\begin{equation}
 H_{{\small \vl}} = J_KSa_0^2 \sum_r \vLambda_r 
\cdot (
{\mbox{\boldmath $\ell$}}^{(A)}_{r} + {\mbox{\boldmath $\ell$}}^{(B)}_{r} ), 
\label{INTERACTION(l)}
\end{equation}
\begin{equation}     
H_{{\small \vn}} = J_KSa_0 \sum_r  (-1)^r \vLambda_r  
\cdot 
({\mbox{\boldmath $n$}}^{(A)}_{r} + {\mbox{\boldmath $n$}}^{(B)}_{r} ),  
\label{INTERACTION(n)}                    
\end{equation}
with
\begin{eqnarray}
\vLambda_r & =& [ \mbox{e}^{2ik_Fra_0} L^{\dagger}_{r, \alpha} R_{r, \beta}
+ \mbox{e}^{-2ik_Fra_0} R^{\dagger}_{r, \alpha}L_{r, \beta} \nonumber \\
& + & L^{\dagger}_{r,\alpha}
L_{r, \beta} + R^{\dagger}_{r,\alpha}R_{r, \beta}] {\mbox{\boldmath $\sigma$}}_{\alpha \beta},
\label{SPINDENSITY}
\end{eqnarray}
measuring the spin density on the stripe.
We have here used the notation 
introduced after Eq.~(\ref{KONDOLATTICE}), implying that $\vl^{(i)}_r \equiv \vl_{r,\, j^{(i)} = 1}$ and 
$\vn^{(i)}_r \equiv \vn_{r, \,j^{(i)} = 1}$. By the assumption that the N\'{e}el order 
directions of the $A$ and $B$ domains are shifted by $\pi$ relative to each other it follows 
that $\vn^{(A)}_r = -  \vn^{(B)}_r$, and thus $H_{{\small \vn}}$ vanishes. It is here important 
to realize that there is no relative staggering  of the ferromagnetic $A$ and $B$ components in 
(\ref{INTERACTION(l)}). These rapidly fluctuating fields are independent, with no 
correlations across the stripe. This leaves us with $H_{{\small \vl}}$, which in the 
continuum limit, using (\ref{spindensity}) with $\vJ \equiv \vJ_L + \vJ_R$, takes the form  
\begin{equation}  \label{CONTINUUMINTERACTION(l)}
H_{{\small \vl}}  \rightarrow \, 2J_KSa_0 \int dx \vJ(x) \cdot
\left( {\mbox{\boldmath $\ell$}}^{(A)}(x) + {\mbox{\boldmath $\ell$}}^{(B)}(x)\right).  
\end{equation}
We have here dropped the nonchiral terms that mix $L$- and $R$-fields since these are washed out
by the rapid phase oscillations in (\ref{SPINDENSITY}).

The fact that the stripe electrons couple manifestly only to the fast ferromagnetic 
components of the localized spins $-$ independent of whether the stripe is metallic 
(Hubbard band away from half-filling) or insulating (half-filled band) \cite{FOOTNOTE2} $-$ 
does not mean that the N\'{e}el order parameter dynamics is completely decoupled 
from the stripe electrons. The N\'{e}el field re-enters the problem via the orthogonality 
condition $\vn \cdot \vl = 0$ which constrains the ferromagnetic component to a plane 
that follows its slow and smooth fluctuations. As we shall see next, part of the interaction 
in (\ref{CONTINUUMINTERACTION(l)}) can be absorbed as an effective spin 
density interaction 
among the stripe electrons. Since at low energies the N\'{e}el order direction is 
essentially constant over large patches in Euclidean space-time, this interaction will 
effectively be pinned in spin space, and hence break the spin rotational invariance of the
electron spin dynamics on the relevant time- and length scales.  
This symmetry-breaking effect, driven by the N\'{e}el order in the environment, will 
dramatically influence the correlations of the stripe electrons. 

For simplicity, we now treat the $A$ and $B$ domains separately. Starting with $A$, 
and collecting all terms in the action containing the $\vl^{(A)}-$field defined on the 
$j^{(A)} = 1$ spin array that couples to the stripe, we find from 
(\ref{SPINACTION}) and (\ref{CONTINUUMINTERACTION(l)}) the contribution to 
the partition function:
\begin{equation}
Z_{\small \vl}^{(A)} = \int {\cal D}[ \vl^{(A)} ] \, \mbox{e}^{-{\cal S}[ {\small \vl}^{(A)}]} 
 \, ,
\label{lPARTITIONFUNCTION}
\end{equation}
with
\begin{widetext}
\begin{equation}
{\cal S}[\vl^{(A)} ] = \int_0^{\infty} d\tau \int dx \, \left[ 4J_Ha_0S^2 (\vl^{(A)})^2 +  
2J_KSa_0 (\vJ_L + \vJ_R ) \cdot \vl^{(A)} - 
iS (\vn^{(A)} \times \partial_{\tau} 
\vn^{(A)} ) 
\cdot \vl^{(A)} \right]. \label{lACTION}
\end{equation}
\end{widetext}
Note that we have again changed to the compact notation $\vn_{j^{(A)} = 1} 
\rightarrow \vn^{(A)}, \,
\vl_{j^{(A)} = 1} \rightarrow \vl^{(A)}$, introduced in Eqs.~(\ref{INTERACTION(l)})
and (\ref{INTERACTION(n)}). 
The integral in  (\ref{lPARTITIONFUNCTION}) is Gaussian and can easily be carried 
out. We obtain 
\begin{multline}
\int {\cal D}[ \vl^{(A)} ] \, \mbox{exp}\left(-\!\int d\tau dx \Big[ (\vl^{(A)})^T \vGamma 
\vl^{(A)}\! + \!\vomega  \vl^{(A)}\Big] \right) \\
=  \mbox{exp}\left(\frac{1}{4}\int d\tau dx (\vomega)^T 
{\vGamma}^{-1} \vomega \right)  
\equiv \mbox{exp}(-{\cal S}^{eff}_{{\small \vl}^{(A)}}),   \label{HUBBARDSTRATONOVICH}
\end{multline}
where $\vGamma = 4J_H a_0 S^2 {\bf 1}, \  \vomega = (2J_KSa_0) \vJ_{\perp} - iS 
(\vn^{(A)} \times \partial_{\tau} \vn^{(A)} )$. We have here defined 
$\vJ_{\perp} \equiv \vJ - (\vJ \cdot \vn^{(A)}) \vn^{(A)}$, with $\vJ \equiv \vJ_L 
+ \vJ_R$, as the piece of the electron spin 
density that $-$ via the constraint $\vn^{(A)} \cdot \vl^{(A)} = 0$ $-$ survives the projection 
onto $\vl^{(A)}$. Thus, from (\ref{HUBBARDSTRATONOVICH}) we obtain the 
effective action coming from fluctuations in $\vl^{(A)}$:    
\begin{widetext}
\be
{\cal S}^{eff}_{{\small \vl}^{(A)}} = \int_0^{\infty} \!d\tau \!\int \!dx \,
\left[  - \frac{a_0J_K^2}{4 J_H} \vJ_{\perp} \cdot \vJ_{\perp} + \frac{iJ_K}{8\pi J_H}
({\vn}^{(A)} \!\times \! \partial_{\tau} {\vn}^{(A)}) \cdot \vJ_{\perp}  
  + \frac{1}{16J_H a_0} 
(\vn^{(A)} \! \times \! 
\partial_{\tau} \vn^{(A)})^2 \right].  \label{lEFFECTIVEACTION}
\ee
\end{widetext}
Let us in turn discuss the different contributions to ${\cal
S}^{eff}_{{\small \vl}^{(A)}}$: \\ \\ 
{\em First term in (\ref{lEFFECTIVEACTION})}. \ The first term is an anisotropic spin 
interaction  among the stripe electrons, induced by the coupling to the N\'{e}el-ordered spins 
in the  $A-$domain. As we have already noted, this interaction follows 
the slow 
fluctuations of the $\vn^{(A)}$-field along the stripe, with $\vJ_{\perp}$ constrained to a 
plane orthogonal to $\vn^{(A)}$. To make progress we shall treat $\vn^{(A)}$ in a mean-field 
formulation, and take it to be in a fixed (but arbitrary) direction $\cf{\tilde{\vn}}$, defined 
by the antiferromagnetic order in the $A$-domain. Introducing a coordinate  system $(x, y, z)$ 
with $\hat{z}$ in the direction of $\cf{\tilde{\vn}}$, and using the operator identity 
$J^z_{L/R}J^z_{L/R} = \frac{1}{3}\vJ_{L/R}\cdot\vJ_{L/R}$, valid for chiral bilinears, the first term in 
(\ref{lEFFECTIVEACTION}) is then seen to add the interaction
\begin{equation}
H_{int} = -\frac{a_0J_K^2}{2J_H} \int dx \,  
(\no{J_L^z J_L^z} + \no{J_R^z J_R^z} + J_L^x J_R^x + J_L^y J_R^y ) 
\label{INTADD}
\end{equation}
to the spin-sector stripe Hamiltonian in (\ref{HUBBARDCONTINUUMspin}). The terms diagonal 
in $L$ and $R$ in (\ref{INTADD}) are forward scattering terms which 
renormalize 
the effective spin velocity $\tilde{v}_{s}$ on the stripe, $\tilde{v}_{s}
\rightarrow \tilde{v}^{(A)}_s = \tilde{v}_s - a_0J_K^2/(4\pi J_H)$, while the terms mixing $L$- and $R$-fields describe backscattering of 
electrons and hence, when added to  (\ref{HUBBARDCONTINUUMspin}), shifts the corresponding coupling: 
$\tilde{g}_{0\perp} \rightarrow \tilde{g}_{\perp} + a_0J^2_K/(4 \pi \tilde{v}^{(A)}_s J_H)$. 
\\ \\
{\em Second term in (\ref{lEFFECTIVEACTION})}. \ Let us first recall \cite{TEHJ} that 
$\vn^{(A)} \times \partial_{\tau} \vn^{(A)}$ is an angular momentum density,
which, at the extremum of the action in (\ref{SPINACTION}) is
locked to the ferromagnetic component: $\vn^{(A)} \! \times \!
\partial_{\tau} \vn^{(A)} \sim \vl^{(A)}$. Since $\vl^{(A)}$ 
is a rapidly fluctuating field,
the second term in  (\ref{lEFFECTIVEACTION}), being a pure phase, for
this case averages to
zero already on finite patches in Euclidean space-time, 
and will be ignored in 
the low-energy limit considered here. This amounts to neglect
fluctuations away from the cluster of paths that dominate the
action (\ref{SPINACTION}) for the localized spins when decoupled
from the stripe. As $J_K \ll J_H$, we do not expect these paths
to change much when inserting the stripe, and the argument
applies also in the presence of the stripe. We shall discuss the
limitations of this mean-field type argument below. \\ \\
{\em Third term in (\ref{lEFFECTIVEACTION})}. \ \ The last term in (\ref{lEFFECTIVEACTION}), 
containing only the N\'{e}el field and its time derivative, should be assembled with the 
spin action in (\ref{SPINACTION}). Then, integrating out {\em all} $\vl_{j^{(A)}}$-
fields from (\ref{PARTITIONFUNCTION}) - in exact analogy with the one-dimensional 
treatment of $\vl_{j^{(A)} = 1}$ in (\ref{HUBBARDSTRATONOVICH}) - taking a continuum limit in 
the $y$-direction, and using $( \vn \times \partial_{\tau} \vn)^2 = (\partial_{\tau} \vn)^2$, 
one obtains the effective action for the order parameter field in the $A$-domain:
\begin{eqnarray}\label{NONLINEARSIGMA}
{\cal S} [\vn ] & = & \frac{1}{2{g}_{0}} \int_{0}^{\infty } d\tau \int_{-\infty}^{\infty} dx 
\int_{0}^{\infty} dy \, 
\left[c{(\nabla {\bf n})}^{2}+\frac{1}{c}{(\frac{\partial{\bf n}}{\partial\tau})}^{2}\right] \nonumber \\
& + & S_{phase}[{\bf n}],           
\end{eqnarray}
where the first term is the action for a nonlinear $\sigma$ model (NL$\sigma$M) in a 
semi-infinite plane, with 
parameters $g_0 ^{-1} = S/\sqrt{8} a_0, c = \sqrt{8}JSa_0$ \cite{TEHJ}. 
One piece of the original sum over Berry phases in (\ref{SPINACTION}) 
has been 
absorbed in the NL$\sigma$M,  while the part (\ref{BERRYSUM}) containing 
the N\'{e}el field only, is left as a 
global phase $S_{phase}[{\vn}]$ in (\ref{NONLINEARSIGMA}). 
This phase is an alternating sum over the solid angles $\Phi[\vn({\small \vR}_i, \tau)]$ 
swept by the local $\vn({\small \vR}_i, \tau)$-fields as $\tau$ goes from 0 to $\infty$.
As long as there are no disordering or finite-size effects causing discontinuities in the 
N\'{e}el field, $S_{phase}[{\vn}]$ will be averaged out \cite{Haldane}. For this reason we will 
ignore it for the moment. For the more realistic case of a finite-width antiferromagnetic domain, 
modeled, say, by a spin ladder \cite{SierraReview}, $S_{phase}[{\vn}]$ will come into play, 
requiring a more careful analysis. We shall return to this important issue in
Sec. III.D. \\ \\ 

The analysis carried out for the $A$-domain above can be repeated step by step for
the $B$-domain, and the fluctuations in $\vl^{(B)}$ are seen to give a contribution identical 
to that in (\ref{lEFFECTIVEACTION}), with the index $"A"$ replaced by  $"B"$.
Summing the contributions from the two domains, it follows that
the stripe electrons get described
by an {\em effective low-energy} Hamiltonian
\begin{equation}\label{STRIPEEFFECTIVE}
H_{stripe} = H_{c} + H_{s},
\end{equation}
with $H_c$ defined in Eq.~(\ref{HUBBARDCONTINUUMcharge}), 
while 
\begin{widetext}
\begin{equation}\label{STRIPEspinEFFECTIVE}
H_{s} = 2 \pi v_{s}\int \, \mbox{d}x \Big\{\no{J^z_{L}
J^z_{L}} + 
\no{J^z_{R} J^z_{R}} - g_{0s} J^{z}_{L}J^{z}_{R} - 
g_{0\perp}(J_{L}^{x} J_{R}^{x} + J_{L}^{y} J_{R}^{y})\Big\},
\end{equation}
\end{widetext}
where
\begin{eqnarray}
g_{0s} & = & \frac{a_0}{\pi v_F}(U-2V), \nonumber \\
g_{0\perp} &=&  \frac{a_0}{\pi v_F} (U - 2V +
\frac{J_K^2}{2J_H} ), \label{RENORMCOUPLINGCOLLINEAR} \\
v_s & = & \tilde{v}_s - \frac{a_0}{2\pi}\frac{J_K^2}{J_H}, \nonumber   
\end{eqnarray}
with $\tilde{v}_s$ defined in Eq.~(\ref{velocities}).

Thus, in contrast to the charge degrees of freedom which remain untouched by the 
coupling to the antiferromagnetic environment, the spin dynamics on the 
stripe is strongly 
renormalized by this same interaction and gets controlled by an {\em effective 
$U(1)$-symmetric} spin Hamiltonian (\ref{STRIPEspinEFFECTIVE}). As for the
low-energy processes in the decoupled antiferromagnetic environment, these are described 
by two independent NL$\sigma$ models, one for each domain $i = A, B$, as defined in 
({\ref{NONLINEARSIGMA}). (It may be worthwhile pointing out that as long as the N\'{e}el field is protected 
by the low-energy thermodynamic limit, the phase $S_{phase}[{\vn}]$ in (\ref{NONLINEARSIGMA}) 
remains inactive \cite{Haldane}.) 

\subsection{Diagonal Stripes}

The construction of the low-energy theory for a stripe running along the diagonal of the 
square lattice (FIG. 2) closely parallels that for a collinear stripe in the previous section. 

\begin{figure}[tbh]
\begin{center}
\includegraphics[width=0.4\textwidth]{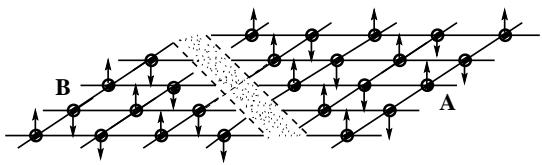}
\caption[DIAGONAL]{Diagonal stripe structure.}
\label{FIG2}
\end{center}
\end{figure}

We again model the isolated stripe by an extended $U-V$ Hubbard
chain, but with a new
hopping matrix element $t' < t$ (since the overlaps between tight-binding orbitals along 
the diagonal of a lattice plaquette is expected to be smaller than along the bonds). As a 
result, the continuum theory in (\ref{HUBBARDCONTINUUMcharge}) and (\ref{HUBBARDCONTINUUMspin})
still applies, but with $t$ replaced by $t'$ (implying a shift of the effective velocities 
defined in (\ref{velocities})).
As for the antiferromagnetic domains, we expect that the order 
parameter dynamics is still described by the NL$\sigma$M  
(\ref{NONLINEARSIGMA}) {\em in the bulk} (away from the stripe): 
By assumption, 
the coupling to the stripe is weak, $J_K \ll J_H$, and can only perturb spins 
in its immediate neighborhood. Here we are interested in the reverse effect, and will 
explore how the Kondo lattice interaction (\ref{KONDOLATTICE}) affects the 
electron dynamics on the diagonal stripe. 

For this purpose, let us isolate one array of localized spins adjacent to the stripe, say  
the $A$-domain. We label the corresponding spin operators $\vS^{(A)}_r$, 
where $r$ is a lattice coordinate running along the stripe that labels also the horizontal 
lattice axes that pierce the stripe at the corresponding sites.


The $\vS^{(A)}_r$ spins interact with their neighboring spins on the parallel array, 
$\vS^{(\tilde{A})}_r$ call them, via the terms
\begin{equation}
H_{array} = J_H \sum_j ( \vS^{(A)}_r \cdot \vS^{(\tilde{A})}_r + \vS^{(A)}_{r-1} \cdot 
\vS^{(\tilde{A})}_r ).   \label{ARRAYINT}
\end{equation}
It is clear from the geometry that this interaction will generate an effective ferromagnetic 
coupling between the nearest-neighbor $\vS^{(A)}$-spins, induced by a double-exchange via the 
two neighboring $\vS^{(\tilde{A})}$-spins. As seen in FIG. 2, this means that the local 1D 
magnetic environment sampled by the stripe electrons via the Kondo exchange is 
{\em ferromagnetically} ordered.  Does this imply a different induced interaction among the 
stripe electrons as compared to the collinear case in (\ref{STRIPEspinEFFECTIVE})? To find out, 
let us first write down the full Kondo lattice interaction for the diagonal stripe, including 
the neighboring array of $\vS^{(B)}_r$-spins from the $B$-domain: 
\begin{multline}
 \label{DIAGKONDOLATTICE}
H_{Kondo} \\ = J_K\!\sum_{r, \, \alpha} c_{r,\alpha}^{\dagger}\vsigma_{\alpha \beta}
c_{r, \beta} 
\cdot ( \vS^{(A)}_{r} + \vS^{(A)}_{r+1} + \vS^{(B)}_{r} + \vS^{(B)}_{r-1}), \\ J_K > 0.
\end{multline}
Following the same route as in Sec II.A.3, using the decompositions 
(\ref{LINEARIZATION}) and (\ref{HALDANEANSATZ}), we obtain:
\begin{equation}
H_{Kondo} = H_{{\scriptstyle \vl}}  + H_{{\scriptstyle \vn}}, 
\label{KONDOLATTICE2}
\end{equation}
where
\begin{equation}
H_{{\scriptstyle \vl}} = J_K S a_0^2 \sum_r \vLambda_r \cdot 
( \vl^{(A)}_r + \vl^{(A)}_{r+1}  + \vl^{(B)}_r +  \vl^{(B)}_{r-1} ),
\label{DIAG(l)}
\end{equation}
\begin{equation}
H_{{\scriptstyle \vn}} = J_K Sa_0 \sum_r \vLambda_r \cdot 
( \vn^{(A)}_r + \vn^{(A)}_{r+1}  + \vn^{(B)}_r +  \vn^{(B)}_{r-1} ),
\label{DIAG(n)}
\end{equation}
with
$\vLambda_r$ the local spin density operator on the stripe defined in (\ref{SPINDENSITY}). 
Similar to the collinear stripe treated above, the $\pi$-phase shifted N\'{e}el order across the 
diagonal stripe - along horizontal as well as vertical directions - implies that 
$ \vn^{(A)}_r = - \vn^{(B)}_r, 
\vn^{(A)}_{r+1} = - \vn^{(B)}_{r-1}$ (see FIG. 2), and it follows that 
$ \mbox{H}_{{\scriptstyle \vn}}$ in (\ref{DIAG(n)}) vanishes, {\em independent of the 
value of the stripe band filling $n_e/2$} \cite{FOOTNOTE2}.  
(Note that, in contrast to the collinear case in (\ref{INTERACTION(n)}),
there is no staggering factor in Eq.~(\ref{DIAG(n)}): the local
magnetic environment as seen from the stripe is uniform {\em
along} the stripe.) We are thus 
left with  $\mbox{H}_{{\scriptstyle \vl}}$ in (\ref{DIAG(l)}). Taking a continuum limit 
\begin{equation}
a_0 \sum_r \rightarrow \int dx\, dy \,\delta(x-y) \, ,
\label{rCONTINUUM}
\end{equation}
and doing a gradient expansion to ${\cal O}(a_0^2)$, we obtain
\begin{multline}
H_{{\scriptstyle \vl}} \rightarrow \\ 4J_KSa_0 \int dx      
 \, \vJ(x,x) 
\cdot \left(\vl^{(A)}(x,x) + \vl^{(B)}(x,x)\right), \\
\vJ(x,x) \equiv \vJ_L(r) + \vJ_R(r), 
\label{STRIPECONTINUUM}
\end{multline}
with the "diagonal" stripe coordinate $r$ replacing 
(the implicitly defined variable) $x$ in the definition of $\vJ_{L/R}$ after Eq.~ 
(\ref{CONTINUUMINTERACTION(l)}). Given (\ref{STRIPECONTINUUM}), we now 
again focus on the $A$-domain, and isolate the piece of (\ref{STRIPECONTINUUM}) 
containing only the $\vl^{(A)}$-field, $H_{\vl^{(A)}}$ call it. Inserting a time-dependence 
and letting $H_{\vl^{(A)}}$ act at (imaginary) time slice $\tau$, integrating over the slices, 
and adding the result to the ordinary semi-classical spin action ${\cal S}_A[\vn, \vl]$ for 
the $A$-domain (cf. with  Eq.~(\ref{ACTION}) for the corresponding collinear case), we 
obtain 
\begin{widetext}
\begin{multline}
{\cal S}_A[\vn, \vl] + \int_{0}^{\infty} d\tau H_{\vl}  \\
 =  \int_{-\infty}^{\infty} dx \int_{x}^{\infty} dy  
\int_{0}^{\infty} d\tau \, \left[ \frac{J_HS^2}{2} \left({(\nabla {\bf n}^{(A)})}^{2} + 
8 (\vl^{(A)})^2 - iS (\vn^{(A)} \times 
\partial_{\tau} \vn^{(A)}) \cdot \vl^{(A)} \right)
+ 4J_KSa_0 \vJ \cdot \vl^{(A)} \, \delta(x-y) \right].
\label{DIAGACTION}
\end{multline}
\end{widetext}
Writing down the full 
$A$-domain action in (\ref{DIAGACTION}) we have ignored the presence of the sum 
over Berry phases (cf. again Eq.~(\ref{ACTION}) for the corresponding collinear case), 
since it does not involve the $\vl$-field and hence does not couple directly to the stripe 
electrons. Also note that compared to our treatment of the collinear case in Sec. II.A.3 
we have here shortcut the analysis by taking a continuum limit in the $y$-direction  
{\em before} integrating out the $\vl$-field in (\ref{DIAGACTION}). Carrying
out the integration 
we obtain an effective $A$-domain action ${\cal S}^{eff}_{{\vl}^{(A)}}$ 
generated by fluctuations in the ferromagnetic components of the localized spins:
\begin{widetext}
\begin{equation}
{\cal S}^{eff}_{{\vl}^{(A)}} = \int_{-\infty}^{\infty} dx \int_{x}^{\infty} dy  \int_{0}^{\infty} d\tau 
\left[ \left( - \frac{a_0J_K^2}{J_H} \vJ_{\perp} \cdot \vJ_{\perp} + 
\frac{i J_K}{2 J_H} (\vn \times \partial_{\tau}\vn) \cdot \vJ_{\perp} \right) \, \delta(x-y) 
 + \frac{1}{16J_H} (\vn \times \partial_{\tau} \vn)^2 \right],  \label{DIAGEFFACTION}
\end{equation}
\end{widetext}
with $\vJ_{\perp}$ defined after Eq.~(\ref{HUBBARDSTRATONOVICH}). 
Doing a saddle-point approximation and dropping 
the rapidly oscillating phase in (\ref{DIAGEFFACTION}), 
we obtain $-$ in exact 
analogy with the collinear case $-$ the induced stripe-electron interaction
\begin{equation}
H_{int} = -\frac{2 a_0J_K^2}{J_H} \int dr \,  (J_L^x J_R^x + J_L^y J_R^y + 
J_L^z J_L^z + J_R^z J_R^z ), \label{DIAGINTADD}
\end{equation}
with $r$ the "diagonal" coordinate along the stripe.

This Hamiltonian is almost an exact copy of that for a collinear stripe, cf. Eq.~(\ref{INTADD}),
with the only difference that the magnitude of the coupling is larger by a factor of 4. 
(This can  be traced back to the fact that {\em on the lattice}
a diagonal stripe electron couples 
simultaneously to \em two} localized spins in the $A$-domain.) 
Adding the identical contribution from the $B$-domain, it follows that the low-energy 
spin dynamics on a diagonal stripe is described by the {\em same} effective 
Hamiltonian (\ref{STRIPEspinEFFECTIVE}) as for the collinear case, but with the renormalized 
parameters in (\ref{RENORMCOUPLINGCOLLINEAR}) modified by 
$J_K \rightarrow 2J_K$.
(As mentioned above, the hopping matrix element $t'$ is also different from that which 
enters the free part of the Hamiltonian 
for a collinear stripe; cf. Eq.~(\ref{STRIPEEFFECTIVE}) with $v_F = 2a_0t\mbox{sin}(\pi n_e/2)$. 
However, as this has no impact 
on the problem studied here, from now on we use a single label $t$ for both types of stripes.)

In contrast to our analysis for the collinear stripe configuration, we have here not 
attempted  to derive the full effective spin action for the decoupled environment. 
As we have already noted, away from the stripe the N\'{e}el order parameter 
dynamics is described by a NL$\sigma$M with an added Berry phase, as in 
(\ref{NONLINEARSIGMA}). By inspection one finds that close to the diagonal stripe 
the Berry phase gets influenced by the unusual boundary condition associated with the diagonal stripe orientation.
Thus, our results - here derived for a semi-infinite 2D geometry
- may be of limited applicability for the case of  ''diagonal''
finite-width or ''spin ladder'' environments (see Sec. III.D). Their study is an
interesting problem, but we here leave it for the future.

The fact that the same effective interaction appears for the diagonal 
and collinear stripe structures (up to the trivial $J_K \leftrightarrow
2J_K$ shift) reflects its origin in the coupling of the electronic
spin density to the {\em uniform $\vl$-components of the
localized spins}. These components are confined to a plane
orthogonal to the $\Neel$-direction, and are blind to whether the
local $\Neel$-field adjacent to a stripe is staggered (as for a
collinear stripe) or uniform (diagonal stripe) \cite{Langmann}. 
The coupling constant $J_K^2/J_H$ embodies the second-order
process that drives the induced interaction between
the stripe electrons: An electron exchanges spin with the
environment $(\sim J_K)$ and another electron arriving at the same lattice
site flips back the localized spin by a second exchange $(\sim
J_K)$, resulting in an effective spin exchange $(\sim J_K^2)$ 
between the two electrons. Since only that part of the spin
exchange that couples to the $\vl$-components survives, 
the effective interaction becomes anisotropic. The $1/J_H$-
dependence of the process is also expected: The larger the spin
stiffness of the antiferromagnetic environment, the smaller the
probability for the double exchange to occur.  \\ \\


\subsection{Effective Stripe Hamiltonian: Bosonization}

As we have seen, the low-energy electron dynamics on collinear as well as 
diagonal 
stripes - taking into account a weak spin exchange with the environment -  
is described 
by an effective Hamiltonian (\ref{STRIPEEFFECTIVE}), with the amplitudes for 
forward- and backward scattering renormalized by the exchange. 
This Hamiltonian 
embodies a {\em spin-anisotropic interaction} among the electrons, well-defined on length- and time 
scales over which the environment is magnetically ordered. To  analyze the 
consequences for the stripe electron dynamics we shall use 
Abelian bosonization to map the model onto two  independent 
quantum sine-Gordon models (in the weak-coupling
limit)- one describing the collective charge excitations, 
the other the spin excitations - and then perform a renormalization group analysis to 
identify the leading instabilities of the system.  

The method of bosonization is well reviewed in the literature \cite{Fradkin,GNT,Senechal}, and we 
here only sketch the most important steps so as to fix notation and conventions. The standard 
bosonization formulas for spinful chiral electrons are given by \cite{Senechal}:
\begin{eqnarray}
R_{\alpha}(x) & = & \frac{1}{\sqrt{2\pi a_0}} \eta_{\alpha} \mbox{e}^{-i\sqrt{4\pi} 
\phi_{R,\alpha}(x)}, \nonumber \\
R_{\alpha}^{\dagger}(x) & = & \frac{1}{\sqrt{2\pi a_0}} \eta_{\alpha} 
\mbox{e}^{i\sqrt{4\pi}\phi_{R, \alpha}(x)}, \label{BD1}
\end{eqnarray}
\begin{eqnarray}
L_{\alpha}(x) & = & \frac{1}{\sqrt{2\pi a_0}} \bar{\eta}_{\alpha} \mbox{e}^{i\sqrt{4\pi}
\phi_{L,\alpha}(x)}, \nonumber \\
L_{\alpha}^{\dagger}(x) & = & \frac{1}{\sqrt{2\pi a_0}} \bar{\eta}_{\alpha} \mbox{e}^{-i\sqrt{4\pi}
\phi_{L,\alpha}(x)}, \label{BD2}
\end{eqnarray}
Here $\phi_{R,\alpha}(x)$ and $\phi_{L,\alpha}(x)$ are right- and left moving bosonic 
fields respectively, carrying spin $\alpha = \, \uparrow,
\downarrow$. The Klein factors $\eta_{\alpha}$ 
and  $\bar{\eta}_{\alpha}$ are inserted to ensure that the anticommutation relations for 
electron fields with different spin come out right. They are Hermitian and 
satisfy a Clifford algebra
\begin{equation}
\{\eta_{\alpha}, \eta_{\beta} \} = \{\bar{\eta}_{\alpha}, \bar{\eta}_{\beta} \} = 
2 \delta_{\alpha \beta}\, ,  \ \ \ \ \ \ 
\{\eta_{\alpha}, \bar{\eta}_{\beta} \} = 0  . \\   \label{KLEIN}
\end{equation}

One next introduces {\em charge} $(c)$  and 
{\em spin} $(s)$ fields $\varphi_{c,s}$ and their duals $\vartheta_{c,s}$:
\bea
\varphi_{c}&=&(\phi_{\uparrow}+ \phi_{\downarrow})/\sqrt{2}, \qquad \vartheta_{c}=
(\theta_{\uparrow} +\theta_{\downarrow})/\sqrt{2}, \\
\varphi_{s}&=&(\phi_{\uparrow}-\phi_{\downarrow})/\sqrt{2}, \qquad \vartheta_{s}=
(\theta_{\uparrow} -\theta_{\downarrow})/\sqrt{2},  \label{BOSONS} 
\eea
where
\be
\phi_{\alpha}=\phi_{L,\alpha}+\phi_{R,\alpha} \, , \qquad
\theta_{\alpha}=
\phi_{L,\alpha}-\phi_{R,\alpha}.
\ee
Then, using the identities
\bea
J_{R}+J_{L}&=&-(1/\sqrt{\pi})\partial_{x}\varphi_{c}\, , \nonumber \\
J_{R}-J_{L}&=&-(1/\sqrt{\pi})\partial_{x}\vartheta_{c}\, , \label{BD3} 
\eea
\bea
J_{R}^z + J_{L}^z &=&-(1/\sqrt{2\pi})\partial_{x}\varphi_{s}\, , \nonumber \\ 
J_{R}^z - J_{L}^z &=&-(1/\sqrt{2\pi})\partial_{x}\vartheta_{s}\, , \label{BD4}
\eea
together with (\ref{BD1}) and (\ref{BD2}),
we can translate $H_{stripe} = H_{c} + H_{s}$ in (\ref{STRIPEEFFECTIVE}) into bosonized form:
\begin{widetext}
\begin{equation}
H_{c} = \frac{v_{c}}{2} \int dx\Big\{\, (\partial_{x}\varphi_{c}^{\prime})^2 +
(\partial_x \vartheta_{c}^{\prime})^2 + \delta_{{\small 1} n_e} 
\frac{2m_{c}}{a_0^2} \kappa\cos(\sqrt{8\pi K_{c}}\varphi_{c}^{\prime})\Big\}, 
\label{SGc}
\end{equation}
\begin{equation}
H_{s} = \frac{v_{s}}{2} \int dx\Big\{\, (\partial_{x}\varphi_{s}^{\prime})^2
+ (\partial_x\vartheta_{s}^{\prime})^2 + 
\frac{2m_{s}}{a_0^2}\kappa\cos(\sqrt{8\pi K_{s}}\varphi_{s}^{\prime})\Big\}.  
\label{SGs} 
\end{equation}
\end{widetext}
We have here introduced the rescaled charge and spin fields 
\begin{equation}
\varphi^{\prime}_{c, s} = K^{-1/2}_{c, s} \varphi_{c, s}, \qquad
\vartheta^{\prime}_{c, s} = K^{1/2}_{c, s}\vartheta_{c, s},
\label{RESCALEDBOSONS}
\end{equation}
and the short-hand $\kappa
\equiv \eta_{\uparrow} \eta_{\downarrow}\bar{\eta}_{\uparrow}
\bar{\eta}_{\downarrow}$.
To leading order in the coupling constants the sine-Gordon model 
parameters $K_{c (s)}$ and $m_{c (s)}$ are given by
\begin{eqnarray}
2(K_{c}-1) & = & g_{0c} = - \frac{a_0}{\pi v_F}(U+6V), \nonumber \\
2\pi m_{c} & = & - g_{0u} = - \frac{a_0}{\pi v_F}(U-2V),  
\label{Kc} 
\end{eqnarray}
\begin{eqnarray}
2(K_{s}-1) & = & g_{0s} = \frac{a_0}{\pi v_F}(U-2V), \nonumber \\
2\pi m_{s} & = & g_{0\perp} = \frac{a_0}{\pi v_F}(U-2V+
\beta\frac{J^2_K}{J_H}), 
\label{Ks}
\end{eqnarray}
with
\begin{eqnarray}
v_c &=& v_F + \frac{a_0}{2\pi}(U+6V),  \\
v_s &=& v_F - \frac{a_0}{2\pi}(U-2V+2\beta\frac{J_K^2}{J_H}),
\label{v}
\end{eqnarray} 
where
\begin{equation}
\beta =\left\{
\begin{array}{l}
1/2 \hskip0.4cm \mbox{collinear stripe} \\
2 \hskip0.5cm \mbox{diagonal stripe}
\end{array}
\right. \, .  \label{BETAPARAMETER}
\end{equation}
Note that in obtaining (\ref{SGc}) and (\ref{SGs}), terms corresponding to scattering processes 
which lead to a renormalization of the Fermi velocities in second order in the couplings, as well as 
strongly irrelevant terms 
$\sim \cos(\sqrt{8\pi K_{c}}\varphi_{c}^{\prime})
\cos(\sqrt{8\pi K_{s}}\varphi_{s}^{\prime})$
describing Umklapp 
processes with parallel spins, have been omitted. 

The product of Klein factors in (\ref{SGc}) 
and (\ref{SGs}) acts on a Hilbert space different from the boson Hilbert space and 
introduces a certain ambiguity into the formalism. We resolve it by choosing a representation of 
the Clifford algebra in terms of tensor products of Pauli matrices and the identity operator 
\cite{Senechal},
\begin{eqnarray}
\eta_{\uparrow} = \sigma_1 \otimes \sigma_1\, , &  \qquad &  \eta_{\downarrow} = \sigma_3 \otimes \sigma_1 \, , 
\nonumber \\
\bar{\eta}_{\uparrow} = \sigma_2 \otimes \sigma_1 \, , & \qquad  &  \bar{\eta}_{\downarrow} = 
\1 \otimes \sigma_2\, , \label{KLEINREP}
\end{eqnarray}
in which the above product $\kappa = \eta_{\uparrow}
\eta_{\downarrow} \bar{\eta}_{\uparrow} \bar{\eta}_{\downarrow}$
of Klein factors has the form
\begin{equation}
\kappa  = 
\1 \otimes \sigma_3 \, .\label{KLEINPRODUCT}
\end{equation}
This matrix is diagonal with eigenvalues $\pm 1$. Provided that all relevant correlation 
functions to be calculated contain only products of Klein factors which are simultaneously 
diagonal with $\kappa$ we 
can pick the eigenstate with eigenvalue $+1$, say, and then ignore the rest of the Klein Hilbert 
space (allowing us to do the replacement $\kappa 
\rightarrow 1$ in (\ref{SGc}) and (\ref{SGs})). We will come back to this 
point below.

\section{Pairing and Density Correlations}

\subsection{Renormalization Group Analysis}

The mapping of the effective stripe Hamiltonian in Eq.~(\ref{STRIPEEFFECTIVE}) onto 
the quantum theory of two independent charge and spin Bose fields, manifestly shows that 
the collective low-energy charge-  and spin dynamics on the stripe remains separated in 
the presence of a magnetic environment.   
This allows us to extract the ground state properties of the
stripe electrons by performing  independent renormalization group (RG) analyses
of the charge- and spin sector
sine-Gordon Hamiltonians. The RG flows are of Kosterlitz-Thouless type, with
effective coupling constants $g_i \, (i=c, s, u, \perp ),$ governed by the 
equations \cite{Wiegmann}
\begin{eqnarray}
dg_{c}/d\ell & = & - g^{2}_{u}, \nonumber\\
dg_{u}/d\ell & = & - g_{c} g_{u} ,
\label{RGCHARGE}
\end{eqnarray}
for the charge sector, and
\begin{eqnarray}
dg_{s}/d\ell & = & - g^{2}_{\perp},\nonumber\\
dg_{\perp}/d\ell & =& - g_{s} g_{\perp},
\label{RGSPIN}
\end{eqnarray}
for the spin sector. 
Here $\ell = \ell n(a/a_{0})$ with $a$ a
renormalized length, while $g_i(\ell =0) \equiv g_{0i}$ are the bare
parameters that enter Eqs.~(\ref{Kc})
and (\ref{Ks}). We shall denote by $\tilde{K}_{c (s)}$ and 
$\tilde{m}_{c (s)}$ the corresponding {\em renormalized} 
sine-Gordon parameters connected to $g_i$ via the same Eqs.~(\ref{Kc})
and (\ref{Ks}).

The flow lines lie on the hyperbolas
\begin{equation}  \label{flowlines}
g^2_{c(s)} - g^2_{u(\perp)} = g^2_{0 c(s)} - g^2_{0 u(\perp)}\, ,
\end{equation}
and $-$ depending on the relation between $g_{0 c(s)}$ and $g_{0
u(\perp)}$ (or, equivalently, the bare sine-Gordon parameters 
$K_{c(s)}$ and $m_{c(s)}$) $-$
exhibit two types of behaviors (cf. FIG. 3): 

\begin{figure}[tbh]
\begin{center}
\includegraphics[width=0.4\textwidth]{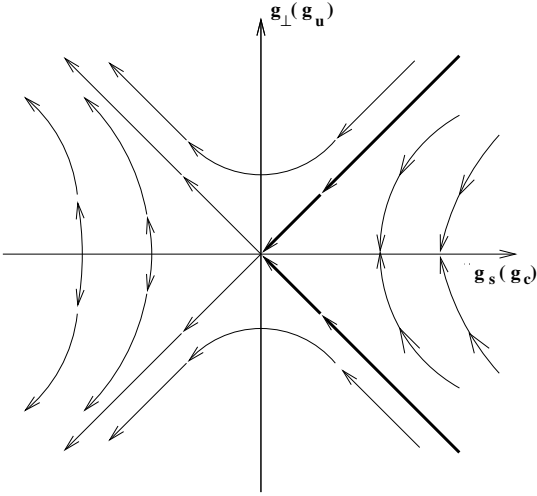}
\caption[RGfig]{Renormalization-group flow diagram for the spin
$(g_s, g_{\perp})$ and charge $(g_c, g_u)$ sectors. The arrows denote
the direction of flow with increasing length scale.}
\label{FIG3}
\end{center}
\end{figure}

{\em Weak coupling regime.} \\ When $ g_{0 c(s)} \ge \left| g_{0
u(\perp)} \right| \, (K_{c (s)}-1 \geq \pi \left|m_{c (s)} \right|)$    
we are in the weak coupling {\em (Luttinger liquid)} regime: 
$g_{u(\perp)} \rightarrow 0$, implying that
the renormalized masses $\tilde{m}_{c (s)}$ scale to zero. 
The low-energy, long-wavelength behavior of  the gapless charge (spin) degrees of 
freedom is thus described by a free scalar field 
\begin{equation}  
H_{c (s)} = \frac{v_{c (s)}}{2}  \int dx \left\{ (\partial_{x} \varphi^{\prime}_{c (s)})^{2}
+ (\partial_{x} \vartheta^{\prime}_{c (s)})^{2} \right\}.  \label{LUTTINGERLIQUID}
\end{equation}
Ignoring logarithmic corrections \cite{GiamarchiSchulzPRB} coming from the slow renormalization of marginally 
irrelevant operators near the fixed line $g_{u(\perp)} = 0$, the
large-distance behaviors of the charge- and spin field correlators and their duals are given by
\begin{equation}
\langle \, \mbox{e}^{i\sqrt{2\pi K^{\ast}_{c(s)}}\varphi^{\prime}_{c (s)}(x)} \, 
\mbox{e}^{-i \sqrt{2\pi K^{\ast}_{c(s)}} 
\varphi^{\prime}_{c (s)}
(0)} \, \rangle \sim \left| x \right|^{- K^{\ast}_{c(s)}},
\label{FREECORRELATIONS1} 
\end{equation}
\begin{equation}
\langle \, \mbox{e}^{i \sqrt{2\pi/ K^{\ast}_{c(s)}}\vartheta^{\prime}_{c (s)}(x)} 
\, \mbox{e}^{-i\sqrt{2\pi/ K^{\ast}_{c(s)}}\vartheta^{\prime}_{c (s)}(0)}
\, \rangle \sim \left| x \right| ^{-1 / K^{\ast}_{c(s)}}.
\label{FREECORRELATIONS2}
\end{equation}
Hence, the only parameters  
controlling the low-energy behavior in the gapless
regimes are the fixed-point values $K^{\ast}_{c(s)}$ 
{\em (Luttinger liquid parameters)} 
of the renormalized coupling constants $\tilde{K}_{c(s)} \approx
1 + g_{c(s)}/2$.  \\ \\ \\
{\em Strong coupling regimes.} \\ When $g_{0 c(s)} < \left| g_{0
u(\perp)} \right| \, (K_{c (s)}-1< \pi \left|m_{c (s)} \right|)$ 
the system scales 
to strong coupling. The two
separatrices $g_{0 c(s)} = \pm \left| g_{0 u(\perp)} \right|$ divide
the strong coupling regimes for charge and spin into two sectors
respectively:
(i) $g_{0 c(s)} \le - \left| g_{0 u(\perp)} \right|$, where the
increase of $\left| g_{c(s)} \right|$ and $\left| g_{u(\perp)} \right|$
is immediate, and (ii) $- \left| g_{0 u(\perp)} \right| < g_{0
c(s)} < \left| g_{0 u(\perp)} \right|$, where one observes a {\em
crossover} from a weak-coupling behavior at intermediate scales
$(g_{c(s)} \approx \left| g_{u(\perp)} \right|)$ to strong coupling
at larger scales $(g_{c(s)} \approx - \left| g_{u(\perp)} \right|)$
\cite{GNT}.
Depending on the sign of the bare mass $m_{c(s)}$ in (\ref{Kc}) and
(\ref{Ks}), the 
renormalized mass $\tilde{m}_{c(s)}$ is driven to $\pm\infty$, signaling a  flow to one of the two 
strong coupling regimes, with a dynamical generation of a commensurability gap $\Delta_{c (s)}$ 
in the charge (spin) excitation spectrum.
The flow of $\left| \tilde{m}_{c (s)} \right|$ to large values 
indicates that the $\mbox{cos}$-term in the 
sine-Gordon model dominates the large-distance properties of the charge (spin) sector. With the 
cosine-term being the dominant one, the values of $\varphi^{\prime}_{c(s)}$ will tend to be 
pinned at the minima of the cosine potential. For $m_{c (s)} < 0$ these are at 
$\sqrt{8\pi K_{c (s)}} \varphi^{\prime}_{c(s)} = 2\pi n$, with $n$ an arbitrary integer. 
Since $\varphi^{\prime}_{c(s)}$ are angular variables one cannot distinguish between different 
$n$, however, and the ``negative mass'' condensation is defined by 
$\langle\varphi^{\prime}_{c(s)}\rangle = 0$. Similarly, for $m_{c (s)} > 0$ the minima are 
at $\sqrt{8\pi K_{c (s)}} \varphi^{\prime}_{c(s)} = \pi n$ and the fields order at 
$\sqrt{\pi / 8K_{c (s)}}$. To summarize, there are two strong-coupling regimes where the 
fields $\varphi^{\prime}_{c (s)}$ get ordered with the expectation values 
\begin{equation}  \label{ORDERFIELDS}
\langle\varphi^{\prime}_{c(s)}\rangle =\left\{ 
\begin{array}{l}
\sqrt{\pi/8K_{c (s)}}, \hskip0.5cm m_{c(s)}>0 \\ 
0, \hskip2.0cm m_{c(s)}<0
\end{array}
\right. \, .
\end{equation}
Note that the signs of the bare masses in (\ref{ORDERFIELDS}) are contingent upon the choice of 
truncated Klein Hilbert space in Sec. II.C, where we have taken 
$\kappa \rightarrow 1$
in (\ref{SGc}) and (\ref{SGs}). This has no effect on the physics, however, 
since a transposition of the two strong-coupling phases above (via the alternative choice 
$\kappa \rightarrow - 1$) 
would be followed by a subsequent redefinition of any relevant correlation 
function, thus producing the same value of any observable.  \\

Having exposed the properties of the weak- and strong coupling regimes, let us apply the
results first to the {\em charge sector}. By inspection of the ``bare'' values of the 
coupling constants in the charge sector, Eq.~(\ref{Kc}), 
one easily finds, using Eqs.~(\ref{ORDERFIELDS}), that for a {\em half-filled band} $(n_e\!=\!1)$  
this sector is gapped for 
$U > 2|V|$ {\em and} for $U < 2V$ when $V>0$ {\em (strong coupling regimes)}.
In the former case $\tilde{m}_{c} \rightarrow -\infty $, 
implying that  
\begin{equation}  \label{PHIc1}
\langle \varphi^{\prime}_{c} \rangle = 0,
\end{equation}
while in the latter case  $\tilde{m}_{c} \rightarrow \infty $, with 
\begin{equation}  \label{PHIc2}
\langle\varphi^{\prime}_{c}\rangle =\sqrt{\displaystyle{\pi/8K_{c}}}.
\end{equation}

For any other values of $U$ and $V$, but still at half-filling, we are in the
{\em weak-coupling regime}, corresponding to a gapless charge excitation spectrum. 
The charge degrees of freedom are here governed by the free Bose field in 
Eq.~(\ref{LUTTINGERLIQUID}), with the fixed-point value of the charge 
parameter 
\begin{equation} \label{Kfix}
K^{\ast}_{c} \simeq 1 + \frac{2}{\pi v_{c}}\sqrt{V(U+2V)} >1, \ \ \ \ n_e = 1,
\end{equation}
obtained from (\ref{flowlines}) with $\tilde{m}_c = 0$.
The line $U=2V>0$ corresponding to the fixed-point line $m_{c}=0, K_{c}-1<0$, is special. 
Here the low-energy properties of the gapless charge sector are described by the free 
massless Bose field in (\ref{LUTTINGERLIQUID}) with $K^{\ast}_c = K_c$,
reflecting its exact marginality.  

{\em Away from half-filling} ($n_e \neq 1$) the bare mass term in (\ref{SGc}) is 
killed off for any values of $U$ and $V$, and the charge degrees of freedom are 
described by the free Bose field  
in (\ref{LUTTINGERLIQUID}). Analogous to the special line $U=2V>0$ above, the correlations in
(\ref{FREECORRELATIONS1}) and (\ref{FREECORRELATIONS2}) are now governed by the bare value of
the Luttinger liquid charge 
parameter in (\ref{Kc}): 
\begin{equation} \label{Kbare}
K^{\ast}_c = K_c = 1 - \frac{a_0}{2\pi v_c}(U+6V), \ \ \ \ n_e \neq 1.
\end{equation}

All of the above is familiar from  
conventional  "g-ology" for Hubbard-type models \cite{Voit,Emery}. Since the  Kondo lattice 
interaction, Eq.~(\ref{KONDOLATTICE}), does not couple to the charge sector the latter indeed 
behaves as if the electrons were isolated from the magnetic environment. \\ 

Let us now look at the behavior of the {\em spin sector}, which is more interesting. As we have seen, the spin exchange between 
the N\'{e}el-ordered environment and the stripe electrons breaks the $SU(2)$ spin-rotational 
symmetry in the effective theory. This implies that the spin sector is gapped for arbitrary 
$U \neq 2V-\beta J_K^2/J_H$ {\em (strong coupling regime)}. 
When $U> 2V-\beta J_K^2/J_H$ the mass renormalization goes to
$+ \infty$, whereas for  $U< 2V-\beta J_K^2/J_H$ the mass renormalizes to 
$- \infty$. Reading off from (\ref{ORDERFIELDS}), using (\ref{Ks}), 
this implies the spin field orderings 
\begin{equation}  
\langle\varphi^{\prime}_{s}\rangle =\left\{ 
\begin{array}{l}
\sqrt{\pi/8K_{s}}, \hskip0.5cm U > 2V - \beta J_K^2/J_H \\ 
0, \hskip1.7cm U < 2V - \beta J_K^2/J_H
\end{array}
\right. \, .  \label{SPINORDER}
\end{equation}
Note that this result is {\em independent} of the band filling on the stripe.
 
When $J_K =0$  and the stripe decouples from the environment, the SU(2) invariance of the 
spin sector is recovered, and the spin dynamics renormalizes along the separatrix $g_s = 
g_{\perp}$.  As is well-known, 
the spin sector then gets controlled by the {\em weak-coupling} Luttinger liquid parameter 
$K_s \simeq 1+{\frac{1}{2}}g_{s} \rightarrow K_s^{\ast} = 1$ when $U > 2V$ \cite{Emery}, 
whereas 
for $U < 2V$ one stays in the strong coupling regime with $\langle\varphi^{\prime}_{s}\rangle = 0$ 
\cite{Voit2,GJ1}.
  
Next, we want to exploit the RG results derived above to map out the groundstate phase diagram 
for the stripe electrons. We shall catalogue the different phases according to the values of 
$(U, V, J_K^2/J_H, n_e)$, and focus on the corresponding 
behaviors of density and superconducting pairing fluctuations. These are characterized 
by the correlations of the associated order parameters, which in the present case come 
in two guises: (1) {\em conventional} and (2) {\em composite} order parameters. 
Let us in turn review their definitions and bosonized representations.

\subsection{Order Parameters}

\subsubsection{Conventional Order Parameters}

The conventional order parameters \cite{Emery,GNT,GJ1,GJ2} which may develop long-range 
correlations in this class of models are those of short wavelength $(2k_{F}x)$ fluctuations of
the {\em site-} and {\em bond-located charge 
density, site-} and {\em bond-located spin density} and {\em superconducting singlet} and 
{\em triplet pairing}. By using the chiral decomposition (\ref{LINEARIZATION}) of the 
electron fields together with the bosonization dictionary in Eq.~(\ref{BD1}) - 
(\ref{BD4}) it is straightforward to obtain their bosonized forms.

\begin{itemize}
\item {\em Charge density wave.}
\end{itemize} 
This fluctuation is carried by the {\em charge-0, spin-0} excitations created by the operator
\begin{multline}  \label{CDW}
{\cal O}_{CDW}(r)  = \sum_{\alpha} c_{r, \alpha}^{\dagger} c_{r, \alpha} \\ 
 = \sum_{\alpha} \! \Big(J_{R,\alpha}(r)\!+\!J_{L,\alpha}(r)\!+\! 
\delta_rL_{\alpha}^{\dagger}(r) R_{\alpha}(r)\!+\! 
\delta_r^{-1}R_{\alpha}^{\dagger}(r) L_{\alpha}(r)\Big)  
\end{multline}
with $\delta_r = \mbox{e}^{2ik_Fra_0}$, and where $J_{L/R}$ are the chiral charge currents defined
in Eq.~(\ref{chargedensity}).
Keeping only the finite-momentum modes $(k = \pm 2k_F$) from the nonchiral terms, taking a 
continuum limit, and reading off from the dictionary (\ref{BD1})-(\ref{BD4}), 
one obtains the bosonized expression
\begin{multline}
{\cal O}_{{\small CDW}}(x) \rightarrow \\
\sin (\sqrt{2\pi K_{c}}\varphi^{\prime}_{c}(x) - 2k_F x) 
\cos (\sqrt{2\pi K_{s}}\varphi^{\prime}_{s}(x)), \label{bosCDW}
\end{multline}
where we have used that $\bar{\eta}_{\uparrow} \eta_{\uparrow}$ and $\bar{\eta}_{\downarrow} \eta_{\downarrow}$ 
are diagonal with the same eigenvalue on the truncated Klein Hilbert space chosen in 
Sec II.C.

\begin{itemize}
\item {\em Spin-density wave.} 
\end{itemize} 
This is the simplest {\em charge-0, spin-1} vector order parameter, and is defined by
\begin{equation}
\vO_{{\small SDW}}(r) = \frac{1}{2} c^{\dagger}_{r, \alpha} 
\vsigma_{\alpha \beta} c_{r, \beta}.
\label{SDW}
\end{equation}
Bosonizing the $x$-component of $\vO(n)_{{\small SDW}}$ that creates $k= \pm 2k_F$ excitations, 
and dropping the trivial $k=0$ modes, one finds in the long-wave length continuum limit
\begin{multline}
{\cal O}^x_{{\small SDW}}(x) \rightarrow \\ \bar{\eta}_{\uparrow} \eta_{\downarrow}
\cos (\sqrt{2\pi K_{c}}\varphi^{\prime}_{c}(x) - 2k_Fx) \sin 
(\sqrt{2\pi K^{-1}_{s}}\vartheta^{\prime}_{s}(x)).
\label{bosSDWx}
\end{multline}
To obtain this form we have exploited the fact that the Klein factors 
$\bar{\eta}_{\uparrow}{\eta}_{\downarrow}$,
$-\bar{\eta}_{\downarrow}{\eta}_{\uparrow}$,
${\eta}_{\uparrow}\bar{\eta}_{\downarrow}$ and $-{\eta}_{\downarrow}\bar{\eta}_{\uparrow}$ have 
the same action in the truncated Klein Hilbert space. (Note, however, that 
$\bar{\eta}_{\uparrow} \eta_{\downarrow}$ is not diagonal on this space, and as a reminder 
of this we keep it explicitly in Eq.~(\ref{bosSDWx})). 
In the same way one easily obtains
\begin{multline}
{\cal O}^y_{{\small SDW}}(x) \rightarrow \\ \bar{\eta}_{\uparrow} \eta_{\downarrow}
\cos (\sqrt{2\pi K_{c}}\varphi^{\prime}_{c}(x) - 2k_Fx)
\cos(\sqrt{2\pi K^{-1}_{s}}\vartheta^{\prime}_{s}(x)),
\label{bosSDWy}
\end{multline}
and
\begin{multline}
{\cal O}^z_{{\small SDW}}(x) \rightarrow \\
\cos (\sqrt{2\pi K_{c}}\varphi^{\prime}_{c}(x) - 2k_Fx)
\sin(\sqrt{2\pi K_{s}}\varphi^{\prime}_{s}(x)),
\label{bosSDWz}
\end{multline}
where in the $z$-component we have put $\bar{\eta}_{\uparrow}\eta_{\uparrow}
= \bar{\eta}_{\downarrow}\eta_{\downarrow} =$ constant. 

In the special case of a half-filled band ($n_e=1$) one can distinguish between the $2k_{F}$ 
modulations of the charge and spin densities with extrema of the 
density profile located {\em on} 
sites or {\em between} sites - i.e. on bonds. Therefore at half-filling one should also consider order 
parameters corresponding to the short wavelength fluctuations of {\em bond-located} charge and 
spin densities:

\begin{itemize}
\item {\em Bond-located charge density wave (``dimer'').}  \
\end{itemize}

A dimerization instability is characterized by enhanced correlations among the 
{\em charge-0, spin-0} excitations created by
\begin{equation}
{\cal O}_{\small bCDW}(r) = \sum_{\alpha} ( c_{r, \alpha}^{\dagger} c_{r+1, \alpha}
+ h.c. ).
\label{DIMER}
\end{equation}
Again, keeping only the $k = \pm 2k_F$ excitations, one obtains in the continuum limit
\begin{multline}
{\cal O}_{\small bCDW}(x) \rightarrow \\ \cos (\sqrt{2\pi K_{c}}\varphi^{\prime}_{c}(x) -
2k_Fx) \cos (\sqrt{2\pi K_{s}}\varphi^{\prime}_{s}(x)).
\label{bosDIMER}
\end{multline}

\begin{itemize}
\item {\em Bond-located spin-density wave.} \
\end{itemize}

This is the vector order parameter that describes {\em charge-0, spin-1}
magnetic excitations centered on the lattice bonds:
\begin{equation}
\vO_{{\small bSDW}}(r) = \frac{1}{2}\sum_{\alpha, \beta}
( c_{r,\alpha}^{\dagger} \sigma_{\alpha \beta} c_{r+1, \beta} + h.c.).
\label{BONDSDW}
\end{equation}
In the continuum limit the lattice shift in (\ref{BONDSDW}) shows up
as an extra phase $\pi/2$ added to the ubiquitous phase $2k_Fx$ 
(cf. the bosonized dimer operator in (\ref{bosDIMER}))
and we thus identify the bosonized components of the 
finite-momentum part of $\vO_{{\small bSDW}}(x)$ as 
\begin{widetext}
\be\label{SDWopY}
{\cal O}^{i}_{bSDW}(x) \rightarrow 
\left\{ 
\begin{array}{l}\bar{\eta}_{\uparrow} \eta_{\downarrow} 
\sin \big(\sqrt{2\pi K_{c}}\varphi^{\prime}_{c}(x)- 2k_Fx \big) 
\sin \big(\sqrt{2\pi K^{-1}_{s}}\vartheta^{\prime}_{s}(x)\big),  \ \ \ \, i =x, \\ 
\bar{\eta}_{\uparrow} \eta_{\downarrow} 
\sin \big(\sqrt{2\pi K_{c}}\varphi^{\prime}_{c}(x)- 2k_Fx \big) 
\cos \big(\sqrt{2\pi K^{-1}_{s}}\vartheta^{\prime}_{s}(x)\big), \ \ \   i =y, \\ 
\sin \big(\sqrt{2\pi K_{c}}\varphi^{\prime}_{c}(x)- 2k_Fx \big) 
\sin\big(\sqrt{2\pi K_{s}}\varphi^{\prime}_{s}(x)\big), \ \ \ \ \ \ \ \ \ \ \ \ i =z. 
\end{array}
\right.
\ee

Finally, we consider the two order parameters for
(superconducting) pairing:

\begin{itemize}
\item {\em Singlet pairing.} \
\end{itemize}
The {\em charge-2e, spin-0} superconducting pairing modes on the stripe lattice are created 
by the operator
\begin{equation}
{\cal O}_{SS}(r) = c^{\dagger}_{r,\uparrow} c^{\dagger}_{r,\downarrow} \\
 = \delta_r L_{\uparrow}^{\dagger}(r)L_{\downarrow}^{\dagger}(r) + 
\delta_r^{-1}R_{\uparrow}^{\dagger}(r) R_{\downarrow}^{\dagger}(r) +
L_{\uparrow}^{\dagger}(r) R_{\downarrow}^{\dagger}(r) + 
R_{\uparrow}^{\dagger}(r) L_{\downarrow}^{\dagger}(r). 
\label{SP}
\end{equation}
The $k = \pm 2k_F$ excitations produced by the chiral terms are the so called 
{\large $\eta$}-pairing modes \cite{ETApairing}. The right-moving {\large $\eta$}-pairs
can be written as  
\begin{equation} \label{ETAMODEleft} 
\mbox{{\large $\eta$}}_R(x) \equiv R_{\uparrow}^{\dagger}(x) R_{\downarrow}^{\dagger}(x) 
\rightarrow \mbox{exp}(i \sqrt{2\pi K_c^{-1}} \vartheta_c^{\prime}(x)) \,  
\mbox{exp}(- i ( \sqrt{2\pi K_c} \varphi_c^{\prime}(x) - 2k_F x)) 
\end{equation}
in the long-wave length limit, with the analogous expression for left-moving 
pairs, $\mbox{{\large $\eta$}}_L(x) \equiv L_{\uparrow}^{\dagger}(x) L_{\downarrow}^{\dagger}(x)$. 
As these contain only the charge field 
and its dual, they are blind to the antiferromagnetic environment and hence we will 
not consider them here. This leaves us with the $k=0$ BCS singlet pairing operator contained 
in (\ref{SP}), with the bosonized form
\\
\begin{equation}
{\cal O}_{{\small SS}}(x) \rightarrow \eta_{\uparrow} \bar{\eta}_{\downarrow}
\, \mbox{exp} (i\sqrt{2 \pi K_c^{-1}} \vartheta^{\prime}_c(x) ) \, \mbox{cos}( \sqrt{2 \pi K_s}
\varphi^{\prime}_s(x) ). \label{BCS}   
\end{equation}
\vspace{2.0cm}
\begin{itemize}
\item{\em Triplet pairing.} 
\end{itemize}  
The {\em charge-$2e$, spin-1} pairing modes are created by the lattice operator
\begin{equation}
{\vO}_{TS}(r) = -i c_{r, \alpha}^{\dagger} (\vsigma \sigma^y)_{\alpha \beta} 
c_{r, \beta}^{\dagger}\, . \label{TRIPLET}
\end{equation}
Again retaining only the $k=0$ modes,
\begin{equation}
{\vO}_{TS}(r) \rightarrow -i\,R^{\dagger}_{\alpha}(r) (\vsigma \sigma^y)_{\alpha \beta}
L^{\dagger}_{\beta}(r)\, ,
\end{equation}
we obtain for the bosonized components in the long-wave length limit:
\be\label{TP}
{\cal O}^{i}_{TS} \rightarrow
\left\{ 
\begin{array}{l}
\mbox{exp}\big( i \sqrt{2\pi K^{-1}_c} \vartheta_c^{\prime}(x) \big) \, \mbox{sin} 
\big(\sqrt{2\pi K^{-1}_s} \vartheta_s^{\prime}(x) \big), \ \ \ \ \ \ \! i =x, \\ 
\mbox{exp}\big( i \sqrt{2\pi K^{-1}_c} \vartheta_c^{\prime}(x) \big) \, \mbox{cos} 
\big(\sqrt{2\pi K^{-1}_s} \vartheta_s^{\prime}(x) \big), \ \ \ \ \ i =y, \\ 
\eta_{\uparrow} \bar{\eta}_{\downarrow} \mbox{exp}( i \sqrt{2\pi
K^{-1}_c} 
\vartheta_c^{\prime}(x)) \, \mbox{sin} 
(\sqrt{2\pi K_s} \varphi_s^{\prime}(x) ), \ \ \ i =z, 
\end{array}
\right.
\ee
\end{widetext}

\subsubsection{Composite Order Parameters}

In addition to the conventional order parameters listed above we need to consider {\em composite} 
order parameters built from operators acting on the stripe electrons {\em and} the magnetic 
environment. The notion of composite order parameters was first exploited in the theory of 
superconductivity \cite{CompositeOrder1,CGT}, where it was 
realized that since any product of a particle-hole (i.e. charge-neutral) operator and a Cooper pair 
operator possess charge $2e$ this composite can, in principle, describe some superconducting state. 
By analogy, one may similarly construct composite CDW and SDW order parameters.

We shall here introduce only composite order parameters that may develop long-range 
correlations for the physically most interesting case of a stripe away from half-filling and with 
repulsive electron-electron interactions $U-2V >0$: composite CDW and composite singlet pairing.

\begin{itemize}
\item {\em Composite (site-located) charge density wave.}
\end{itemize}
A composite CDW order parameter 
is obtained by projecting the conventional (site-centered) spin-1/2 SDW onto the difference between 
the localized spins on the neighboring $A$- and $B$-arrays:
\begin{equation}
{\cal O}_{{\small c-CDW}} \sim \vO_{{\small SDW}} \cdot ( \vS^{(A)} - \vS^{(B)} ).
\label{COMPCDW}
\end{equation}
Note that this expression is well-defined in the continuum limit for any stripe geometry: In 
particular, for the case of a diagonal structure, an electron at the $r^{th}$ site on the stripe 
couples to $\vS_{r-1}^{(A)} + \vS_{r}^{(A)}$ in the $A$-domain, which in the continuum limit reduces to 
$2\vS^{(A)}(x_r)$, dropping an irrelevant gradient term. 
Considering first a collinear structure, we need to keep only 
the staggered parts $\vn_r^{(A)}(-1)^r$ and $\vn_r^{(B)}(-1)^r$
of the localized spins since the correlations of the uniform   
components of $\vS_r^{(A)}$ and $\vS_r^{(B)}$ die out fast
\cite{Fradkin}.  
With a phase-antiphase domain, 
as assumed here, we further have that $\vn^{(A)} = - \vn^{(B)} \equiv \vn$. 
Thus, from (\ref{COMPCDW}), the $k=2k_F + \pi/a_0 \equiv
2k_F^{\ast}$ part of the
composite charge density wave
is given by
\begin{equation}
{\cal O}^{(k=2k_F^{\ast})}_{{\small c-CDW}} \sim {\vO}_{SDW} 
\cdot \vn (-1)^r,
\label{COMPCDWeff}
\end{equation}
with $r$ the stripe lattice coordinate, and with the bosonized components of ${\vO}_{SDW}$ written 
down in Eqs.~(\ref{bosSDWx}) - (\ref{bosSDWz}). 
It would be tempting to refer to the {\em generalized Luttinger
theorem} \cite{Yamanaka} to ''explain'' the
appearance of the composite staggered CDW, Eq.~(\ref{COMPCDWeff}). As
pointed out by Zachar \cite{Zachar}, the theorem asserts that
theories belonging to the class of Kondo-Heisenberg lattice type
models are expected to support a massless spin-$0$, charge-$0$
excitation of momentum $k = 2 k_F^{\ast}$ (reflecting the
presence of a ''large Fermi surface'' due to the
localized spins). However, in the present case the localized
spins of the environment are assumed to be ordered (with the NL$\sigma$M in 
(\ref{NONLINEARSIGMA}) describing the small fluctuations of the order parameter field
$\vn$), and, as a consequence, time reversal symmetry - entering
as a condition for the validity of the theorem \cite{Yamanaka} -
is broken. Indeed, the case of a diagonal structure is
different, and does {\em not} produce a $k = 2 k_F^{\ast}$
mode. Here the stripe electrons experience a local ferromagnetic
environment (cf. FIG. 2), and the composite CDW now appears at
$k=2k_F$ (i.e. with {\em no} staggering):
\begin{equation}
{\cal O}^{(k=2k_F)}_{{\small c-CDW}} \sim {\vO}_{SDW}
\cdot \vn \, .
\label{COMPCDWeffdiag}
\end{equation}

\begin{itemize}
\item{\em Composite singlet pairing.} \ \
\end{itemize}
By taking the product of the conventional triplet pairing 
operator $\vO_{TS}$ for the stripe
with the difference of spin operators for localized spins, $\vS^{(A)} - \vS^{(B)}$, a composite
singlet pairing operator can be formed as
\begin{equation}
{\cal O}_{{\small c-SS}} \sim \vO_{TS} \cdot ( \vS^{(A)} - \vS^{(B)} ).
\label{CSP}
\end{equation}

For a collinear stripe this operator has two momentum components: 
a uniform $k=0$ composite singlet
\begin{displaymath}
\vO_{TS} \cdot ( \vl^{(A)} - \vl^{(B)} ) \, ,
\label{CSPUNIFORM}
\end{displaymath}
with rapidly decaying correlations due to the incoherent fluctuations of $\vl^{(A)}$ and
$\vl^{(B)}$, and a $k=\pi/a_0$ staggered composite singlet
\begin{eqnarray} \label{CSPSTAGGERED}
{\cal O}^{(k=\pi/a_0)}_{{\small c-SS}} & \sim & \vO_{TS} \cdot 
\vn (-1)^{r} \\ \nonumber
& = & -i\,R^{\dagger}_{\alpha}(r) (\vsigma \sigma^y)_{\alpha \beta}
L^{\dagger}_{\beta}(r) \cdot
\vn (-1)^{r},
\end{eqnarray}
with $r$ the discrete lattice coordinate along the stripe, and where 
$\vn \equiv \vn^{A} = - \vn^{B}$.   
It is important to note that ${\cal O}^{(k=\pi/a_0)}_{{\small c-SS}}$ 
is {\em odd} under time
reversal ($T: R \leftrightarrow L, \, \vsigma \rightarrow - \vsigma,
\, \vn \rightarrow - \vn$), as well as under 
parity ($P: R \leftrightarrow L$), 
implying ''odd-frequency odd-parity pairing''
\cite{OddFrequency1}. 

Turning to the case of a diagonal stripe
structure the composite singlet pairing now occurs for $k=0$
(since the local magnetic environment appears uniform
as seen from the stripe), and one has
\begin{eqnarray} \label{diagcspstaggered}
{\cal O}^{(k=0)}_{{\small c-SS}} & \sim &
-i\,R^{\dagger}_{\alpha}(r) (\vsigma \sigma^y)_{\alpha \beta}
L^{\dagger}_{\beta}(r) \cdot
\vn \, .
\end{eqnarray}
Again, parity and time reversal are broken. It is important to point out that
theoretical work \cite{ColemanMirandaTsvelik} suggests that odd-frequency pairing
is actually unstable for $k=0$ pairs (at least within Eliashberg-Migdal theory,
where vertex corrections to the self-energy are neglected). This result
becomes particularly
intriguing when seen in the light of the diagonal $\rightarrow$ collinear
stripe rotation associated with the superconducting transition observed in some of
the cuprates \cite{Wakimoto,Matsuda} (cf. our discussion in Sec. I): If the singlet pairing
in the high-$T_c$ materials were of composite nature, the stripe rotation would precisely serve
to stabilize the pairing by shifting the momentum from $k=0$ (diagonal configuration
with unstable pairing) to $k=\pi/a_0$ (collinear configuration with a
stable, staggered
composite pairing mode)!

In the presence of 2D N\'{e}el order, as assumed
here, the $\vn$-field correlations are infinitely ranged in the
groundstate, and 
the ${\cal O}^{k=\pi/a_0}_{{\small c-SS}}$ 
and ${\cal O}^{k=0}_{{\small c-SS}}$ operators may build up 
large-distance correlations that compete effectively with conventional triplet pairing. 
Whether this happens, and what other order parameter 
correlations may develop, will be studied next. 
For an extended discussion of
composite order parameters for 1D correlated electrons, 
we refer the reader to Refs. 26 and 28.

\subsection{Phases}

Equipped with the results in the two previous sections we shall now pinpoint the leading 
groundstate instabilities of the stripe electrons and list the corresponding phases:  

\subsubsection{Half-filled band: $n_e =1$}

The phase diagram consists of five sectors: A, B, C, D1, and D2 (see FIG. 4).
\begin{itemize}
\item{\bf{A phase}: $U >2|V|$.}
\end{itemize}

We include this case merely as an illustration of our formalism,
as a half-filled band (one electron per
site on the stripe) is somewhat special when combined with
dominant repulsive on-site interaction. 
The reason is that 
one here expects to loose the phase-antiphase
configuration as $n_e \rightarrow 1$, and instead recover the undoped
antiferromagnetic state
(with an {\em in-phase} $\Neel$ configuration across the
''stripe''). In work by
Zachar \cite{Zachar2}, based on a stripe t-J model (corresponding to a ''strong-
coupling'' limit $J_K = J_H$ of our lattice model in (\ref{MODEL})), it was suggested that
there is a transition from the phase-antiphase to in-phase $\Neel$ configuration 
already at a band filling $\sim 0.6$ 
(see also Ref. 57). Still, it is instructive
to explore the assumption of a half-filled stripe embedded in a hypothetical
phase-antiphase $\Neel$ background, and explore the consequences. At half-filling the charge
excitation spectrum is gapped $(\Delta_c \neq 0)$ when $U>2|V|$. The stripe is thus
insulating and the ordering of the charge boson with ground state expectation value
$\cf{\varphi^{\prime}_c} = 0$ suppresses the conventional CDW and superconducting correlations,
but leaves behind the SDW and Peierls (dimerized) correlations. Turning to the spin sector,
according to Eq.~(\ref{SPINORDER}) there is a condensation at
$\langle \varphi^{\prime}_s \rangle =\sqrt{\pi/8K_s}$. This kills off the Peierls
correlations, and as the ``in-plane'' SDW$^{x,y}$-correlations are seen to be incoherent we are
left with
\begin{multline}
\langle {\cal O}^{z}_{SDW} (x) {\cal O}^{z}_{SDW} (x^{\prime}) \rangle
\sim \mbox{cos}(2k_F(x-x^{\prime})) \\ \rightarrow (-1)^l \times {\rm constant},
\label{NEEL}
\end{multline}
where in the last step we have reintroduced the discrete stripe coordinate $x=ra_{0}$,
$x^{\prime}=(r+l)a_{0}$. Thus, given the hypothesis of a phase-antiphase spin background
one obtains a {\em long-ranged
antiferromagnetic (N\'{e}el) phase} for the stripe. The energy of this frustrated configuration
grows linear with the length of the stripe and is hence unphysical, as 
anticipated. Note, however, that the actual ''in-phase'' $\Neel$ configuration for $n_e =1$
{\em does} imply a long-ranged $\Neel$ phase for the stripe!

\begin{figure}[tbh]
\begin{center}
\includegraphics[width=0.4\textwidth]{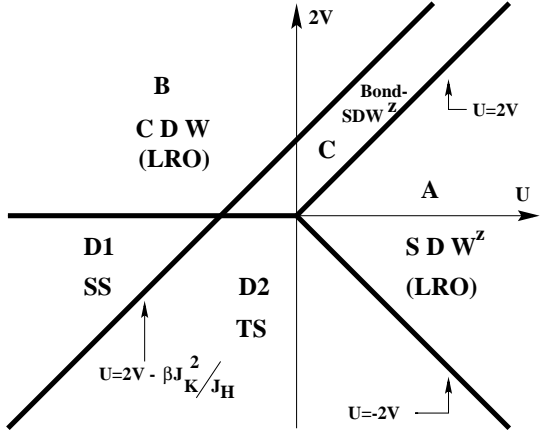}
\caption[Half-filling]{The ground state phase diagram of the stripe electron system at half-filling.
Solid lines separate different phases: SDW$^{z}$ $-$ long-range ordered
spin density wave phase; CDW $-$ long-range ordered charge
density wave phase;
SS $-$ BCS (superconducting) singlet pairing phase; TS$^{z}$ $-$ triplet pairing phase
(coexisting with a composite SS); bSDW$^z$ $-$ long-range ordered bond-located spin density wave.
As explained in the text, the $A$-domain can only be realized with an ''in-phase'' $\Neel$ configuration
across the stripe. The other stripe phases are {\em assumed} to coexist with antiphase $\Neel$ configurations 
of the localized spins.}
\label{FIG4}
\end{center}
\end{figure}

\begin{itemize}
\item {\bf B phase}: $0 < U + \beta J^2_K/J_H < 2V$.
\end{itemize}
This phase is that of an insulator with a long-range ordered CDW: Both charge and spin excitations 
are gapped. The fields $\varphi^{\prime}_{c(s)}$ get ordered with ground state expectation values  
$\langle \varphi^{\prime}_{s}\rangle  = 0$ and 
$\langle \varphi^{\prime}_{c} \rangle =\sqrt{\displaystyle{\pi/8K_{c}}}$ respectively, and  
\begin{equation}
\langle {\cal O}_{\small CDW}(x){\cal O}_{\small CDW}(x^{\prime })\rangle 
\sim (-1)^{l} \times{\rm constant},
\end{equation}
with $l$ defined after Eq.~(\ref{NEEL}).
For the case of an isolated stripe ($J_K\!=\!0$), results from weak-coupling perturbative 
renormalization group studies 
\cite{Emery,Voit2} show that there is a continuous phase transition along the line $U=2V$
separating the SDW$^z$ and CDW phases.
The recent interest in the extended $U-V$ Hubbard 
model was triggered by Nakamura,\cite{Nakamura1}, who found numerical evidence that 
for small to intermediate values of $U$ and $V$, the SDW$^z$ and CDW phases are mediated by  
a bond-located charge-density-wave (bCDW) phase: The SDW$^z$-CDW transition splits into 
two separate transitions: ({\it i}) a Kosterlitz-Thouless spin gap transition from SDW$^z$ to bCDW 
and  ({\it ii}) a continuous transition from bCDW to CDW. An analogous sequence of phase 
transitions in the vicinity of the $U=2V$ line is an intrinsic feature of extended $U-V$ 
Hubbard models with bond-charge coupling \cite{GJ1}. A similar effect is here caused 
by the Kondo coupling to the antiferromagnetic environment: At $J_{K}\neq 0$, along the line 
$U=2V$ only the charge gap closes. Therefore this line correspond to a {\em metallic state} 
with dominating antiferromagnetic SDW$^{z}$ and bSDW$^{z}$ correlations 
(since $K_{c}<1$), showing 
identical power-law decays at large distances:
\begin{eqnarray}
\langle {\cal O}^{z}_{{\small SDW}}(x){\cal O}^{z}_{{\small SDW}}(x^{\prime})\rangle &=& 
\langle {\cal O}^{z}_{{\small bSDW}}(x){\cal O}^{z}_{{\small bSDW}}(x^{\prime})\rangle \nonumber \\ 
&\sim& (-1)^{l}\times |x-x^{\prime}|^{-K_{c}}.
\end{eqnarray}
\\
\begin{itemize}
\item {\bf C phase}: \quad $- \beta J^2_K/J_H < U-2V < 0 $ and $V>0$.
\end{itemize}
Here again a charge gap opens. However, since in this sector $U-2V<0$ the bosonic charge field
is now condensed with ground state  
expectation value $\langle \varphi^{\prime}_{c} \rangle = \sqrt{\pi/8K_c}$. This 
immediately leads to suppression of the site-located SDW$^z$ correlations, and instead the
bond-located bSDW$^{z}$ exhibits long-range order: 
\begin{equation}
\langle {\cal O}^{z}_{{\small bSDW}}(x){\cal O}^{z}_{{\small bSDW}}(x^{\prime})\rangle  
 \sim (-1)^{l} \times {\rm constant}.
\end{equation}
The line $V=0, U<0$ 
is the crossover line from the insulating phases into the superconducting phases. On this 
line the charge sector is in the weak-coupling gapless (metallic) phase with $K^{\ast}_c = 1$. However, 
the spin sector is massive along this line except a the point $U=- \beta J^2_K/J_H$, which marks the 
transition from a metallic phase at $ -\beta J^{2}_{K}/J_{H} < U < 0$ where the 
SDW$^z$, bSDW$^z$ and TS$^z$ fluctuations show identical algebraic decay at large distances
\begin{multline}
\langle {\cal O}^z_{SDW}(x) {\cal O}^z_{SDW}(x^{\prime}) \rangle \sim \langle {\cal O}^z_{bSDW}(x) 
{\cal O}^z_{bSDW}(x^{\prime}) \rangle \\ \sim  \langle {\cal O}^z_{TS}(x) {\cal O}^z_{TS}(x^{\prime}) \rangle
\sim \left| x - x^{\prime} \right|^{-1}\, ,
\label{CORR1}
\end{multline}
to a different metallic phase at $U < -\beta J^{2}_{K}/J_{H}$, where the 
SDW, bSDW and TS$^{z}$ fluctuations are suppressed, 
while the conventional CDW, SS and Peierls correlations show identical large distance behavior:
\begin{multline}
\langle {\cal O}_{CDW}(x) {\cal O}_{CDW}(x^{\prime}) \rangle \sim \langle {\cal O}_{SS}(x) 
{\cal O}_{SS}(x^{\prime}) \rangle \\ \sim  \langle {\cal O}_{dimer}(x) {\cal O}_{dimer}(x^{\prime}) \rangle
\sim \left| x - x^{\prime} \right|^{-1}.
\label{CORR2}
\end{multline}

This large degeneracy of metallic phases along the line $V=0$ is due to the SU(2)-charge 
{\em (``pseudo-spin'')} symmetry of the half-filled Hubbard model \cite{GNT}.
This degeneracy is immediately lifted by an 
attractive nearest-neighbor coupling ($V<0$), in support of superconducting instabilities. One finds two
phases with enhanced pairing correlations:

\begin{itemize}
\item{ {\bf D1 phase}: \quad $ U < 2V - \beta J^2_K/J_H$ and $V < 0$}.
\end{itemize}
Here the dominating instability is towards conventional BCS {\em singlet pairing}, with
correlations 
\begin{equation}
\langle {\cal O}_{SS}(x){\cal O}_{SS}(x^{\prime}) \rangle \sim  \left| x-x^{\prime}\right|^{-1/K_{c}}.
\label{CORRss}
\end{equation}

\begin{itemize}
\item{ {\bf D2 phase}: \quad $2V\! - \!\beta J^2_K/J_H \!< \!U <-2V$, \ $V < 0$}.
\end{itemize}
In this region {\em triplet pairing} shows a power-law decay at large distances
\begin{equation}
\langle {\cal O}^{z}_{TS}(x){\cal O}^{z}_{TS}(x^{\prime}) \rangle \sim  
\left| x-x^{\prime}\right|^{-1/K_{c}},
\label{CORRts}
\end{equation}
and is the dominating instability in the ground state. It follows that the {\em composite singlet
pairing} operator ${\cal O}^{(k=\pi/a_0)}_{{\small c-SS}}$, defined in Eq.~(\ref{CSPSTAGGERED}),
also builds up large-distance correlations:
\begin{eqnarray} \label{CSPcorrelation}
& & \langle {\cal O}^{(k=\pi/a_0)}_{{\small c-SS}}(x) 
{\cal O}^{(k=\pi/a_0)}_{{\small c-SS}}(x^{\prime})\rangle \nonumber \\
 & \sim & (-1)^{\ell} \langle {\cal O}^{z}_{TS}(x){\cal O}^{z}_{TS}(x^{\prime}) \rangle 
\langle n^z(x) n^z(x^{\prime}) \rangle \nonumber \\
& \sim & (-1)^{\ell} \times \left| x-x^{\prime}\right|^{-1/K_{c}}.
\end{eqnarray}
We have here used Eqs.~(\ref{CSPSTAGGERED}) and (\ref{CORRts}), 
together with the property that the $\Neel$ order parameter,
with $\langle \vn(x) \vn(x^{\prime}) \rangle =
\langle n^z(x) n^z(x^{\prime}) \rangle =$ constant, defines the ``out-of-plane'' direction
$\hat{z}$ along which the triplet pairing correlations are enhanced. Similarly, for a diagonal
stripe one would have, using Eqs.~(\ref{diagcspstaggered}) and (\ref{CORRts}):
\begin{equation} \label{CSPdiagcorrelation}
\langle {\cal O}^{(k=0)}_{{\small c-SS}}(x)
{\cal O}^{(k=0)}_{{\small c-SS}}(x^{\prime})
\rangle \sim \left| x-x^{\prime}\right|^{-1/K_{c}}.
\end{equation}
However, as shown by Coleman {\em et al.} \cite{ColemanMirandaTsvelik}, $k=0$ odd-frequency pairing is likely to be
intrinsically unstable, and hence is not expected to compete with the conventional triplet pairing mode.

\subsubsection{Away from half-filling: $n_e \neq 1$}.

We now turn to the physically more relevant case of a stripe with $n_e \neq 1$, 
assuming that $n_e$ is within the range
where the stripe forms an antiphase domain wall between the $\Neel$ 
configurations ($n_e \le 0.6$ in the strong-coupling limit $J_K \sim J_H$, 
according to Refs. 56 and 57).

Away from half-filling the charge sector is always in the {\em (weak coupling)} Luttinger liquid 
metallic regime, controlled 
by the bare value of the Luttinger liquid parameter $K_{c}$. The phase diagram now splits into four
sectors: A, B, D1 and D2 (see FIG. 5), separated by the line $U-2V + \beta J^2_K/J_H =0 $
(where the spin gap closes) and by the crossover line $U+6V=0$ ($K_c =1$) separating the metallic 
phases with dominating density-density correlations at $K_{c}<1$ from those    
with dominating superconducting correlations at $K_{c}>1$. 

When $U-2V + \beta J^2_K/J_H  > 0$ the spin sector flows to strong coupling with the 
excitations condensing at $\langle \varphi^{\prime}_s \rangle = \sqrt{\pi/8K_s}$. This kills off 
the CDW as well as the ``in-plane'' SDW$^{x,y}$-correlations. 
The SDW$^z$ and 
TS$^z$ correlations, on the other hand, survive the spin 
field ordering, and one finds
\begin{eqnarray}
\langle {\cal O}^z_{SDW}(x) {\cal O}^z_{SDW}(x^{\prime}) \rangle 
& \sim & \left| x-x^{\prime}\right|^{-K_c} \label{SDWcorr}, \\
\langle {\cal O}^z_{TS}(x) {\cal O}^z_{TS}(x^{\prime}) \rangle 
& \sim & \left| x-x^{\prime}\right|^{-1/K_c}. \label{TScorr} 
\end{eqnarray}

Therefore, in the 
\begin{itemize}
\item{{\bf A phase}: \quad  $U-2V + \beta J^2_K/J_H > 0; \quad K_{c} <1,$}
\end{itemize}
the SDW$^z$ correlation is the dominating instability in the system (with TS$^z$ subleading), 
while in the  
\begin{itemize}
\item{{\bf D2 phase }: \quad  $U-2V + \beta J^2_K/J_H > 0; \quad K_{c} > 1,$}
\end{itemize}
the {\em triplet pairing} (TS$^z$) fluctuations dominate (with SDW$^z$ being 
subleading). \\

When $U-2V + \beta J^2_K/J_H  < 0$ the spin sector flows to the
other strong coupling regime, with the spin
excitations condensing at $\langle \varphi^{\prime}_s \rangle = 0$, 
the charge 
sector remaining in the weak-coupling metallic region. The correlations that 
exhibit algebraic decay are those of CDW and SS$^z$ excitations, 
and one finds 
\begin{eqnarray}
\langle {\cal O}_{CDW}(x) {\cal O}_{CDW}(x^{\prime}) \rangle 
& \sim & \left| x-x^{\prime}\right|^{-K_c}, \\
\langle {\cal O}^z_{SS}(x) {\cal O}^z_{SS}(x^{\prime}) \rangle 
& \sim & \left| x-x^{\prime}\right|^{-1/K_c}.
\end{eqnarray}
It follows that in the 
\begin{itemize}
\item{{\bf B phase}: \quad  $U-2V + \beta J^2_K/J_H < 0; \quad K_{c} <1,$}
\end{itemize}
the CDW correlation is the dominating instability in the system 
(with SS subleading), 
while in the  
\begin{itemize}
\item{{\bf D1 phase }: \quad  $U-2V + \beta J^2_K/J_H < 0; \quad K_{c} > 1,$}
\end{itemize}
the conventional singlet-pairing BCS fluctuations are the most dominant (with the CDW
fluctuations being subleading).

In the A and D1 phases the composite CDW$^z$ and SS order parameters also build up large-
distance correlations. In exact analogy with the D2 phase at half-filling, one obtains
for a {\em collinear stripe}:
\begin{equation} \label{CCDWcorrelation}
\langle {\cal O}^{(k=2k_F^{\ast})}_{{\small c-CDW}}(x)
{\cal O}^{(k=2k_F^{\ast})}_{{\small c-CDWS}}(x^{\prime})
\rangle  \sim 
(-1)^{\ell}\times \left| x-x^{\prime}\right|^{-K_{c}},
\end{equation}
and
\begin{equation} \label{CSScorrelation}
\langle {\cal O}^{(k=\pi/a_0)}_{{\small c-SS}}(x)
{\cal O}^{(k=\pi/a_0)}_{{\small c-SS}}(x^{\prime})
\rangle  \sim 
(-1)^{\ell}\times \left| x-x^{\prime}\right|^{-1/K_{c}}. 
\end{equation}
For a diagonal stripe a composite CDW also develops, with correlations
\begin{equation} \label{CCDWdiagcorrelation}
\langle {\cal O}^{(k=2k_F)}_{{\small c-CDW}}(x)
{\cal O}^{(k=2k_F)}_{{\small c-CDWS}}(x^{\prime})
\rangle  \sim 
\left| x-x^{\prime}\right|^{-K_{c}}.
\end{equation}
The fate of the zero-momentum composite singlet is more uncertain,
however, considering its intrinsic instability \cite{ColemanMirandaTsvelik}.

\begin{figure}[tbh]
\begin{center}
\includegraphics[width=0.4\textwidth]{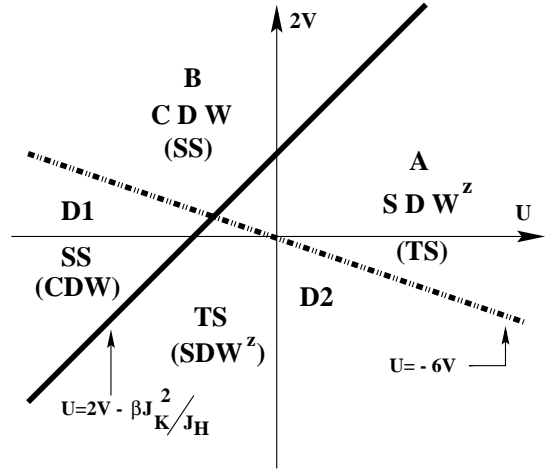}
\caption[Away from half-filling]{The ground state phase diagram of the stripe electron system at $n_{e} \neq 1$.
The solid line corresponds to a narrow metallic phase with gapless spin excitation spectrum, and
separates two
different spin gapped phases: the SDW$^{z}$ and/or TS$^{z}$ phase and the CDW and/or SS phase,
respectively.
The dashed line marks the crossover from a metallic phase with dominating density-density
correlations into a phase with dominating pairing correlations. All instabilities
shown
in the phase diagram exhibit a power law decay of correlations. Subleading instabilities
with correlations which decay faster then the dominating ones are indicated in brackets.
As discussed in the text, the enhanced SDW$^z$ (TS$^z$) correlations coexist with {\em composite}
CDW (SS) instabilities.}
\label{FIG5}
\end{center}
\end{figure}

We summarize our findings in FIG. 5.  We have to stress again the weak-coupling nature of the phase 
diagram. Higher-order corrections will modify the shape of the phase boundaries. However, more 
important are strong coupling effects. In the case of strong values of the Kondo lattice interaction, 
one may expect additional phase transitions due to the finite band width of the effective stripe model. 
Such effects cannot be traced within the continuum limit (infinite band) approach used in this paper, 
and will require numerical studies.


\subsection{The A phase away from half-filling: A scenario for
nonconventional superconductivity?}

Of the phases considered, the A phase away from half-filling is
of particular interest as the conditions $U-2V+\beta J_K^2/J_H
>0; K_c < 1; n_e \neq 1$ are expected to apply to a generic
stripe in a cuprate material: Most experiments \cite{Lavrov}
indicate that the stripes in the cuprates are intrinsically {\em
metallic},  with no commensurability gap even when the stripe
order is static and strong, as in the Nd-doped materials
\cite{Ando}. Although a precise specification of the coupling constants
is beyond present-day technology, the Coulomb interaction among
the stripe electrons is expected to dominate other couplings
(electron-phonon, dopant potentials, interlayer fields,...),
implying the bound $U-2V+\beta J_K^2/J_H >0,$ with $K_c < 1$.
Unfortunately, the electron dynamics on time scales where 
stripe fluctuations can be neglected (for which our model may
apply) is still to be searched out experimentally, and there are
as yet no ''hard data'' against which we can confront
our results. 

The A phase is dominated by a conventional SDW$^z$-instability
together with a composite CDW, 
coexisting with {\em two} subleading superconducting
instabilities: conventional triplet pairing and composite
singlet pairing (breaking parity and time reversal). This is different from
the well-known scenario of a ''spin gap proximity
effect'' \cite{EKZ} where pair-hopping between a stripe and
a {\em spin-gapped} insulating environment ''infects''
the stripe with the gap, resulting in a conventional CDW instability, 
with a subleading singlet pairing channel.
In the case where stripe fluctuations are sufficiently slow that
they can be treated as ''quasi-static'', the CDW instability can be
shown to be suppressed by destructive interference between
neighboring meandering stripes, leaving the singlet
superconducting instability as the leading one
\cite{KFE}.  The singlet order parameter on each
stripe is then assumed to become correlated across the sample via
inter-stripe ''Josephson'' coupling, leading to
superconducting long-range order below a critical temperature.
In contrast, in our scenario singlet
superconductivity (with the added property of {\em breaking parity
and time
reversal symmetry)} would require the suppression of the leading
SDW$^z$
{\em and} (composite) CDW instabilities, in addition to that of 
triplet pairing. 
As we shall see below, quasi-static fluctuations do not perform
this trick. Rather, meandering stripes living on a {\em collinear
backbone} tend to phase lock so that (conventional) {\em triplet
pairing} comes out as the leading effective instability. In the
case of a {\em diagonal structure}, the (composite) singlet
pairing correlations (if at all present; cf. our discussion after Eq.~ 
(\ref{diagcspstaggered}), survive the slow stripe fluctuations,
and coexist with the triplet pairing channel. Whether a complete
theory $-$ treating stripe fluctuations and the one-dimensional
electron dynamics on equal footing $-$ would change our picture in
favor of singlet pairing remains an open question.  

Leaving for future work the problem if and how long-range
superconducting order may emerge from an A-phase type instability
when stripe fluctuations are fully included in the analysis,
there are still several issues that need to be addressed: 
\begin{itemize}
\item{Is the spin gap sufficiently large for the instabilities in
the A phase to survive at finite temperatures?}
\item{What happens when taking into account the fact that the
antiferromagnetic environment as seen by a stripe is {\em not} that
of two semi-infinite domains but is rather made up of two {\em finite-width}
domains, separating the stripe from its neighbors?}
\item{How do transverse stripe fluctuations influence the A-phase
instabilities?}
\item{What about possible long-range interactions among the
stripe electrons?}
\end{itemize}
Let us discuss these questions in turn. \\

{\bf The size of the spin gap.} \
The A phase corresponds to a strong-coupling regime $g_s = - \left|
g_{\perp} \right| $ which is reached after a crossover from weak
coupling (where $g_s = \, \left| g_{\perp} \right| $); cf. Sec. III.A
and FIG. 3.
Because of the crossover, the spin gap opens slowly, 
and it is {\em a priori} not
obvious that it will suffice to sustain the A phase in presence
of thermal fluctuations. To find out, we use (\ref{flowlines}) to
integrate the RG equations (\ref{RGSPIN}), and identify the
length scale where $\left| g_{\perp} \right| $ becomes of order unity.
This scale - where the perturbation is of the same order as the
fixed point Hamiltonian and renders the theory noncritical -
defines the {\em correlation length} $\xi_s$ of the electronic
spin sector. Using $\left| g_{\perp}(\xi_s) \right| \, \sim {\cal O}(1)
\gg \, \left| g_{0s} \right|$ in the integrated scaling equation for
$g_{\perp}$, we obtain
$\xi_s = a_0 \, \mbox{exp}[(\pi/2 - \mbox{arctan}(g_{0s}/\delta g)
)/2\pi \delta g]$, where $\delta g \equiv \sqrt{g_{0\perp}^2 -
g_{0s}^2}$. The associated spin gap $\Delta_s = \hbar v_s \xi_s^{-1}$
is thus given by
\begin{equation} \label{spingap}
\Delta_s = \hbar \frac{v_s}{a_0} \mbox{exp}\Big[ - \frac{\pi/2 -
\mbox{arctan}(g_{0s}/\sqrt{g_{0\perp}^2 - g_{0s}^2} ) }{2\pi 
\sqrt{g_{0\perp}^2 - g_{0s}^2} } \Big],
\end{equation}
with $g_{0s} = a_0(U-2V)/\pi v_s$ and $g_{0\perp} = a_0(U-2V +
\beta J_K^2/J_H)/ \pi v_s$, as given in Eq.~(\ref{Ks}).  There is
considerable latitude in specifying the parameters entering
(\ref{spingap}), but choosing  $v_s \sim 10^5$ m/s, $a_0 \sim$
5\AA \ , $U-2V \sim 10^{-1}$eV, and $J_K^2/J_H \sim 10^{-4}$eV $-$
all within reasonable bounds $-$ we obtain from (\ref{spingap}) a
spin gap $\Delta_s$ corresponding to a temperature of about 500K.
We conclude that the gap is robust, and is
expected to sustain the A phase in the relevant temperature
range
(where stripe formation is possible \cite{TranquadaReview}).
It is interesting to note that an estimate of 500K is about
''right'' if one were to identify the spin gap with
the pseudogap observed in the underdoped metallic phase of the
cuprates \cite{Carlson}. A considerable amount of local pairing would
then be present well beyond the superconducting transition
temperature (in this region determined by the onset of global
phase coherence). \\

{\bf Finite-width antiferromagnetic domains.}  \
A stripe in a real material is not embedded in a 2D
antiferromagnet, but is separated from its neighbors by
finite-width domains, or {\em spin ladders},
with a finite antiferromagnetic correlation length $\xi_{AFM}$.
In order for the A phase to survive in this more harsh
environment the spin gap must develop on a length scale shorter
than $\xi_{AFM}$ (where the stripe electrons can still sample
local magnetic order). This implies the condition
\begin{equation} \label{gapcondition}
\xi_s < \xi_{AFM}\, ,
\end{equation}
with $\xi_s$ the spin correlation length on the stripe. 
The zero-temperature correlation length $\xi_{AFM}$ for
$S=1/2$ spin ladders with an {\em even number of legs} $n_{\ell}$
has been calculated analytically
\cite{Chakravarty}:
\begin{equation} \label{chakravarty}
\xi_{AFM} \approx 0.5 a_0 \mbox{e}^{0.68 n_{\ell}} (1 - 0.73
n_{\ell} ), 
\end{equation}
where, as before, $a_0$ is the lattice spacing. Using that
$\xi_s = \hbar v_s / \Delta_s $, our estimate from above,
$\Delta_s/k_B \sim 500$K, together with (\ref{gapcondition}) and
(\ref{chakravarty}), imply that the A phase survives for even-leg
ladders with $n_{\ell} \ge 4$.  As suggested by the Monte-Carlo
data in \cite{Syljuasen}, a four-leg ladder with $J_H$ of the
order of $10^{-2}$eV may support the A phase up to a temperature
of $T \sim 50$K.  By increasing $U-2V$ and/or $J_K$, the spin gap
grows, allowing for the A phase to persist at the lower
bound $n_{\ell} =4$ for even higher temperatures.

The case of an {\em odd-leg ladder} requires further
analysis. Now, with $S=1/2$, the Berry phase in (\ref{BERRYSUM})
contributes a nonvanishing topological term to the effective
action for the spin ladder, implying a diverging spin-correlation
length $\xi \rightarrow \infty$ but with no antiferromagnetic
order over large distances. 
However, in the weak
coupling regime the topological term is effectively inactive
\cite{Chakravarty}, and as a consequence there is no distinction
between gapless and gapful ladders on length scales shorter than
$\xi_{AFM}$. It follows that the condition (\ref{gapcondition})
is the same for even-leg and odd-leg ladders.
Although we cannot rigorously exclude that nonperturbative
effects may  carry over to the stripe electrons on length scales
{\em larger} than $\xi_{AFM}$, it seems improbable considering
the fact that the spin sector of the stripe develops a mass at a
length scale which is shorter than and independent of $\xi_{AFM}$.
As such the mass is already well-established at the scale where
non-perturbative effects from the ladder may come into play.

Before concluding this discussion we wish to add two more notes:
First, the analysis sketched here is strictly valid only for a
collinear stripe structure. As we commented upon in Sec II.B, the
unusual boundary condition implied by a {\em diagonal stripe
orientation} influences the Berry phase in a nontrivial way. This
may produce a non-negligible feed-back on the localized spins when
considering a finite-width or spin-ladder environment, possibly
changing the A phase in an unexpected way. Secondly,
one should note that the {\em composite order parameters} 
for a collinear structure, defined in
Eqs.~(\ref{COMPCDWeff}) and (\ref{CSPSTAGGERED}), decay faster 
with a spin-ladder
environment as compared to the case of two semi-infinite 2D
domains. It follows that at distances shorter than $\xi_{AFM}$ - where our
construction is still expected to be valid - the algebraic decay of 
the $\Neel$-order
correlations produce a faster decay of the composite
correlations compared to the conventional ones. This is
different from the case where a one-dimensional electron gas is
coupled by a Kondo lattice coupling to two non-interacting
antiferromagnetic Heisenberg spin-1/2 chains (i.e. with $n_{\ell}
=1$). As shown in Ref. 30, the composite order parameters (c-SS
and c-CDW) here induce the dominant instabilities. This reflects
the fact that the model with an $n_{\ell} =1$ environment
renormalizes to a fixed point different from ours, instead
belonging to the universality class of {\em chirally stabilized
liquids} \cite{Andrei}, with {\em no} opening of a spin gap. It is
here important to note that if one were to increase our model
parameters $U-2V$ and/or $J_K$ so that (\ref{gapcondition}) and 
(\ref{chakravarty})
were simultaneously satisfied for 
$n_{\ell} =1$, our construction would break down. In particular,
the assumption of a {\em weak coupling} RG scheme - as exploited
in Sec. III.A - would be violated.
In addition, the pronounced role of the Berry phase for 
$n_{\ell} =1$  is likely to invalidate the construction of the
effective spin sector model
in (\ref{STRIPEspinEFFECTIVE}), as paths in (\ref{SPINACTION}) 
away from the extremum
now enter the stage to influence the ground state also at
short and intermediates length scales. The $n_{\ell} =1$
environment is therefore not expected to be covered by our
approach, and must instead be studied by other methods, such
as that advocated in Ref. 30. \\

{\bf The quasi-static limit: the effect of slow stripe
fluctuations.} \
To study the effect of {\em slow} stripe fluctuations on the
order parameter correlations of phase A we consider a 2D array of
static stripes and take an equilibrium average over their
meanderings. Adopting the notation in Ref. 10,
we use a coordinate system in which
the stripe array runs along the $x$-direction (collinear {\em or}
diagonal on the lattice),
with a transverse displacement of a stripe in the $y$-direction
labeled by $y_j(x)$, with $j$ indexing the stripe. Introducing
the arc length $L_j(x)$, measuring the distance along the
$j^{th}$ stripe to position $x$,
\begin{displaymath}
L_j(x) = \int_ 0^x dx^{\prime} \sqrt{1 +
(\del_{x^{\prime}}y_j(x^{\prime}))^2} \, ,
\end{displaymath}
we infer from Eq.~(\ref{bosSDWz}) the expression for the SDW$^z$ 
order parameter on a meandering stripe:
\begin{equation}  \label{SDWmeander}
{\cal O}_{SDW}^z(j, x) = \tilde{{\cal O}}_{SDW}^z(j, x) + h.c.
\end{equation}
with
\begin{widetext}
\begin{equation}
\label{SDWmeandersplit}
\tilde{{\cal O}}_{SDW}^z(j, x) \\ \sim e^{i(2k_FL_j(x) -\sqrt{2\pi
K_c} \varphi_c^{\prime}(j, x))} \mbox{sin}
\Big[ \sqrt{2\pi K_s} \varphi_s^{\prime}(j, x) \Big]. 
\end{equation}
The coupling of the SDW$^z$ to that of a neighboring stripe is of
the form \cite{SDWUmklapp}
\begin{eqnarray} \label{SDWint}
H_{SDW} &\sim \int dx g[\Delta y_j(x)] & \left( \tilde{{\cal
O}}^{\dagger}_{{SDW}^z}(j, x) \tilde{{\cal O}}_{{SDW}^z}(j\! +\! 1, x)  +
h.c. \right) \nonumber \\ 
  & \sim \int dx g[\Delta y_j(x)] & \mbox{sin}
\Big[ \sqrt{2\pi K_s} \varphi_s^{\prime}(j, x) \Big] \mbox{sin}
\Big[ \sqrt{2\pi K_s} \varphi_s^{\prime}(j\!+\!1, x) \Big]\, 
\mbox{cos} \Big[ \sqrt{2\pi K_c} \Delta \varphi_c^{\prime}(j,
x) - 2k_F \Delta L_j(x)\Big]
\end{eqnarray}
\end{widetext}
where $g( \Delta y_j (x))$ is an $x$-dependent coupling constant.
Neglecting possible overhangs of stripes we have here defined
$\Delta y(j,x) \equiv y(j+1,x) -y(j,x) > 0$ (and similarly for
$\Delta \varphi_c^{\prime}$ and $\Delta L_j$). By integrating out
the stripe fluctuations $y_j(x)$ in powers of $g$
one obtains an effective Hamiltonian
of an equivalent rigid system, with a renormalized coupling
\begin{equation}  \label{effg}
\tilde{g} \sim \langle g[\Delta y_j] \rangle \, \mbox{exp} (-2k_F^2 
\langle (\Delta
L_j )^2 \rangle ) 
+ {\cal O}(g^2),
\end{equation}
where $\langle...\rangle$ denotes an average over meandering
stripes. As argued in Ref. 10,
since the signs of $\Delta L_j(x)$
are randomly distributed along the distance $x$, one expects 
$\Delta L_j(x)$ to grow as a random walk:
$\langle (\Delta L_j(x))^2 \rangle \sim \, \left| x \right| $. 
It follows from
(\ref{effg}) that the interstripe coupling between SDW$^z$'s
can be ignored in the thermodynamic limit.

The conclusion that transverse stripe fluctuations cause
destructive interference between SDW$^z$'s on neighboring
stripes clearly applies to {\em any} $k \neq 0$ order parameter:
The exponential suppression in (\ref{effg}) can be avoided only
if the momentum transfer multiplying $\langle (\Delta L_j (x) )^2 
\rangle $ is
identical to zero.  Thus, the dephasing effect operates also for
the A-phase composite CDW's, Eqs.~(\ref{COMPCDWeff}) and 
(\ref{COMPCDWeffdiag})
(with $k=2k_F + \pi/a_0 \ [k=2k_F]$ for a collinear [diagonal]
stripe backbone), as well as for the staggered $k=\pi/a_0$
composite singlet pairing, Eq.~(\ref{CSPSTAGGERED}), for a collinear
structure \cite{phaseflip}.
In contrast, the conventional ($k=0$) triplet pairing channel
(\ref{TP}) survives the stripe fluctuations.

We caution the reader that the argument, adopted from Ref. 10,
is valid only in the quasi-static
limit where stripe
fluctuations are sufficiently slow to be treated via an annealed
average over (static) meandering stripes. Moreover, the argument
is strictly valid only in the thermodynamic limit. The complete
problem, where the quantum dynamics of {\em mesoscopic} stripes
is treated on equal footing with the intrinsic Luttinger liquid
instabilities remains unsolved. 

For complementary views of the physics of meandering 
and fluctuating stripes we refer the reader to Refs. 78-82. \\ \\

{\bf What about long-range electron-electron interactions?} \
This question becomes critical when realizing that the Coulomb
interaction on an {\em isolated} stripe is poorly screened, given
its insulating environment. Neighboring stripes may provide
metallic screening over a finite range, but our
''assumption'' that this range is of the order of a
lattice spacing - implicitly built into the extended $U-V$ Hubbard
model in (\ref{HUBBARD}) - may not be realistic. Still, having included
a nearest-neighbor repulsion $V $ in the model, we do obtain {\em
some} information about the effect of the  poor  screening: As
can be gathered from Eqs. (\ref{Ks}), (\ref{SDWcorr}) and 
(\ref{TScorr}), the
presence of $V >0$ enhances the SDW$^z$ instability, whereas the
TP$^z$ gets weaker. In the case of an arbitrary finite screening
length $\kappa_s$, Schulz \cite{Schulz} found that the
large-distance correlations are governed by a modified charge
parameter $K_c \sim 1/\sqrt{\mbox{ln} \kappa_s}$.
In the A phase, this again gets translated into stronger
(weaker) SDW$^z$  (TP$^z$) correlations as compared to the case with
only local on-site interactions $\sim U$ (cf. again Eqs. 
(\ref{SDWcorr}) and (\ref{TScorr})) \cite{SchulzFootnote}.
For small $J_K^2/J_H$ (as assumed
here) one expects this result to apply also in the spin-gapped 
A phase. We conclude that as long as there is a finite screening
length $\kappa_s$ present, our results - using the $U-V$ Hubbard
model - should remain at least qualitatively valid for length
scales $>\kappa_s$: The A phase supports an SDW$^z$ instability
with a subleading TS$^z$ (which, however, gets weaker as $\kappa_s$
grows larger). 
Taking into account the finite lengths of the stripes would
introduce yet another scale into the problem (cf. the discussion
above), requiring a more
sophisticated analysis. 

In this context we wish to remind the reader that we obtained the
induced spin-interaction in (\ref{STRIPEspinEFFECTIVE}) by a saddle-point
approximation, with
$\vn \times \del_{\tau} \vn$ in (\ref{lEFFECTIVEACTION}) 
locked to the small but
fast oscillations of the $\vl$-field. Fluctuations away from  the
extremum are expected to produce an effective retarded
interaction in the spin sector, similar to what happens in the
charge sector of a Luttinger liquid when integrating out
electron-phonon interactions \cite{VoitSchulz}. Whereas the {\em
local} fluctuations of the $\Neel$ field could possibly be
treated in analogy with the phonon problem \cite{VoitSchulz}, an
analysis of the dispersive antiferromagnetic excitations (which
may produce a long-range tail of the retarded interaction)
requires a novel theoretical approach. We have to leave this
problem as a challenge for the future.


\section{Twisted Antiferromagnetic Domains}

So far we have have been concerned with the ideal situation of {\em
perfect}
phase-antiphase $\Neel$ configurations surrounding the stripe, 
giving rise to the
effective U(1) symmetric model in (\ref{STRIPEspinEFFECTIVE}). 
In this section we 
generalize the discussion
to the case where fluctuations twist the $\Neel$ configurations 
with respect to each
other, breaking the SU(2) spin-rotational symmetry completely.

How does the $\Neel$ order parameter change across the antiphase
boundary defined by the stripe? Let us take the boundary along $ y = 0 $.
The simplest situation is that
\begin{equation}
\langle\vn (x,y)\rangle  =  f(y) \hat{z}
\end{equation}
with $ f(y) = - f( - y ) $.  This is the situation we have
considered so far;
that the rotational symmetry about the spin $ z $ axis is
preserved, and that the $\Neel$ order parameter simply decreases
across as we approach the antiphase boundary and is reversed on the
opposite side.

A second possibility is that the $\Neel$ order parameter rotates
along (for instance) the spin $ x $ axis as it approaches the
antiphase boundary. In addition, we can let it rotate
around the spin z axis along the boundary, resulting
in the following form for $ \langle \vn(x,y) \rangle$:
\begin{equation} \label{pitch}
\langle \vn(x,y) \rangle = (\cos( \alpha ) \sin( 2 \pi q x ),
         \cos( \alpha ) \cos( 2 \pi q  x ), \sin( \alpha )),
\end{equation}
where $ \alpha = \alpha(y) $ is an odd function
with $ \alpha( \pm \infty ) = \pm \pi/2.$
The parameter $ q $ is the rotational pitch along the antiphase
boundary The special case $ q = 0 $ corresponds to a "collinear" spin
texture, and if $ q \ne 0 $, we find  a phase that is
topological, in the sense that the Neel order parameter  covers
the spin space with topological density of $  4 \pi q $ per unit
length of the phase boundary.

These phases are all further subdivided by their symmetry under
reflections through the antiphase boundary. If the lattice
points are arranged so that the line $ y = 0 $ contains
lattice points, the actual configuration is a configuration under
reflections through $ y = 0 $ and the antiphase boundary is
{\it site centered}. If $ y = \pm 1/2 a_0 $
contains the lattice points, the configuration is even under
reflections through $ y = 0 $ and the configuration is
{\it bond centered.}

These various stripe scenarios have all been investigated as
candidates for spin textures associated with striped antiphase
boundaries. Although the investigations are not conclusive, it
is fair to say that neither experimental nor theoretical investigations
suggest that anything but $q=0$ stripe configurations should be considered
candidates for a spin texture. In fact, simple calculations 
\cite{AnderssonOstlund}
suggest that the spin texture that appears to be energetically favored
is in fact the site centered collinear stripes along $ (10) $  that
are odd under reflections through the antiphase boundary. For the
diagonal $ (11) $ stripes, the bond centered and site centered
have energies that are almost identical. Furthermore, the spin
domains of these stripes are all extremely narrow, as measured
by the width of the functions $ \alpha( y ) $. This supports our
use of an effective 1D model for a site-centered stripe.

To study the effect of a completely broken spin-rotational
symmetry we shall confine our attention to the case
of a collinear stripe (along the $\hat{x}$-direction) 
away from half-filling (cf. sec II.A), and with 
repulsive
interactions among the electrons $(U-2V > 0)$. As before, we denote the two
insulating semi-infinite domains surrounding the stripe by $A$ and $B$ 
respectively.
We may assume that the $\Neel$ directions $\vn^{(A)}(x,y)$ and
$\vn^{(B)}(x,y)$ are parameterized as in Eq.~(\ref{pitch}), with $q=0$
and with $\alpha$ a {\em slowly} varying function of $y$, except across the
stripe where $\alpha$ changes sign. To connect to the local
mean-field picture used in Sec. II.A we shall simply think of two
fixed $\Neel$ directions
$\langle \vn^{(A)}\rangle$ and $\langle \vn^{(B)}\rangle $ which differ by an
{\em arbitrary} angle $\gamma$, $0 < \gamma < \pi$, across the
stripe (with $\gamma = \pi$ for perfect phase-antiphase $\Neel$
configurations).

The construction of the effective model proceeds along the same lines as in Sec II.A
However, since $\vn^{(A)}_r \neq - \vn^{(B)}_r$ when $\gamma \neq \pi$,
we must now pay attention to possible contributions from the coupling $H_{\vn}$
of the lattice spin density to the $\Neel$ order parameters, Eq.~ 
(\ref{INTERACTION(n)}).
By inspection it is easy to verify that the staggered spin density $(-1)^r
\vLambda_r$ entering Eq.~(\ref{INTERACTION(n)}) vanishes in the 
continuum limit {\em
away from
half-filling}, implying that for this case $H_{\vn}$ does {\em not} come 
into play, even when
$\gamma \neq \pi$.
Taking $\langle \vn_r^{(A)}\rangle \ \rightarrow \ 
\langle \vn(x)^{(A)} \rangle \, \sim 
\hat{z}$, as in sec II.A, it
follows from (\ref{lEFFECTIVEACTION}) that the effective 
electron-electron interaction on the stripe
mediated by the Kondo exchange with the {\em $A$-domain} remains the same as before, as 
given in Eq.~(\ref{INTADD}):
\begin{equation}
H^{(A)}_{int} = -\frac{a_0J_K^2}{2J_H} \int dx \,  (\no{J_L^z
J_L^z} + \no{J_R^z J_R^z} + J_L^x J_R^x + J_L^y J_R^y).
\end{equation}
Turning to the $B$-domain, we rotate the coordinate system so that its 
$\Neel$ direction $\langle \vn^{(B)}(x)\rangle$ lies in the $yz$-plane:
\begin{equation}
\langle \vn^{(B)}\rangle \ = \mbox{sin}\gamma\, \hat{y} + \mbox{cos}\gamma\, \hat{z}, 
\ \ \ \ 0 \le \gamma \le
\pi \, ,
\end{equation} 
with $\gamma$ the angle w.r.t. the $\hat{z}$-axis defining the $\Neel$ order direction of the
$A$-domain. The piece of the electron spin density $\vJ$ that survives the projection onto the
plane in which the uniform $\vl^{(B)}$ components live (cf. the discussion in sec II.A) is given by
\begin{eqnarray}
\vJ_{\perp} & \equiv & \vJ - (\vJ \cdot 
\langle \vn^{(B)}\rangle)\langle\vn^{(B)}\rangle
\nonumber \\ 
            & = & \mbox{sin}^2\gamma\, J^yJ^y + \mbox{cos}^2\gamma\,
J^zJ^z \\  
            & + & \mbox{cos}\gamma\,
\mbox{sin}\gamma\, (J^yJ^z + J^zJ^y). \nonumber
 \label{JperpB}
\end{eqnarray}
Decomposing $\vJ \rightarrow \vJ_L + \vJ_R$, and using the chiral identities
\begin{eqnarray} 
J^y_{L/R} J^z_{L/R} &+& J^z_{L/R} J^y_{L/R}  =  0, \nonumber \\ 
J^y_{L/R} J^y_{L/R} &=& J^z_{L/R} J^z_{L/R} \, , \label{chiralcurrents}
\end{eqnarray}
it follows from (\ref{lEFFECTIVEACTION}) and (\ref{JperpB})
that the effective electron-electron stripe interaction mediated 
by the Kondo exchange with the spins in the {\em $B$ domain} is given by
\begin{widetext}
\begin{equation} \label{Bdomain}
H^{(B)}_{int}(\gamma) =  -\frac{a_0J_K^2}{2J_H} \int dx  
\Big\{ \! \no{J^z_LJ^z_L} + \no{J^z_RJ^z_R} + J_L^x J_R^x  
+ \!\mbox{cos}^2\gamma J_L^y J_R^y + \mbox{sin}^2\gamma J_L^z J_R^z  
- \!\mbox{cos}\gamma \,\mbox{sin}\gamma\, (J^y_L J^z_R + J^z_L J^y_R) \Big\}.
\end{equation}
The induced interaction terms (\ref{INTADD}) and (\ref{Bdomain})  
are now to be added
to the spin Hamiltonian $H_s$ in (\ref{HUBBARDCONTINUUMspin}) (which describes the spin sector of
the stripe electrons decoupled from the environment). Writing $H_s + 
H_{int}^{(A)} + H_{int}^{(B)}(\gamma) \equiv H_s(\gamma)$,
we obtain:
\begin{eqnarray}
H_s(\gamma) = 2\pi v_s \int dx &\Big\{& \no{J_L^z J_L^z} + 
\no{J_R^z J_R^z} - 
g_{0s}(\gamma)J_L^z J_R^z 
-  g_{0\perp} J_L^x J_R^x - (g_{0\perp} - \frac{a_0J_K^2}{4\pi v_s J_H}
\mbox{sin}^2\gamma)J_L^y J_R^y 
\nonumber \\  
\  & + & \frac{a_0J_K^2}{4\pi v_s J_H} \mbox{cos}\gamma\,\mbox{sin}\gamma 
\,(J_L^y J_R^z + J_L^z J_R^y)\, \Big\},
\label{twistH}
\end{eqnarray}
with
\begin{equation}
g_{0s}(\gamma) = \frac{a_0}{\pi v_s}(U - 2V + \mbox{sin}^2\gamma\,
\frac{J_K^2}{4J_H}), \label{gs-alpha} 
\end{equation}
and with $g_{0\perp}$ and $v_s$ given in Eq.~(\ref{RENORMCOUPLINGCOLLINEAR}).
Introducing the rotated currents
\begin{equation}
j_{R/L}^x  =  J_{R/L}^x, \ \ \
j_{R/L}^y  =  \mbox{cos}(\theta) J_{R/L}^y + \mbox{sin}(\theta) 
J_{R/L}^z, \ \ \
j_{R/L}^z  =  - \mbox{sin}(\theta) J_{R/L}^y + \mbox{cos}(\theta) 
J_{R/L}^z, 
\label{RotatedCurrents}
\end{equation}
we can write $H_s(\gamma)$ on diagonal form w.r.t.
spin components by choosing $\theta = - \gamma/2$:
\begin{equation}
H_s(\gamma) = 2\pi v_s \int dx \Big\{ \no{j_L^z j_L^z} +
\no{j_R^z j_R^z} - \, g_{0\perp}(j_L^xj_R^x + j_L^yj_R^y)
 -  g_{0s}(\gamma) j_L^zj_R^z + g_{0f}(\gamma) j_L^yj_R^y \Big\},
\label{rotHam}
\end{equation}
\end{widetext}
where
\begin{eqnarray}
g_{0\perp} & = & \frac{a_0}{\pi v_F}(U-2V+\frac{J_K^2}{2J_H}) ,
\nonumber \\
g_{0s}(\gamma) & = & \frac{a_0}{\pi v_F}(U-2V +
\mbox{sin}^2(\gamma/2)\frac{J_K^2}{2J_H}) , \nonumber \\ 
g_{0f}(\gamma) & = & \frac{a_0}{\pi v_F}
\mbox{sin}^2(\gamma/2)\frac{J_K^2}{2J_H} , \nonumber \\ 
v_s & = & v_F - \frac{a_0}{2\pi}(U-2V+\frac{J_K^2}{J_H}) \, .
\label{rotparameters}
\end{eqnarray} 
The four first terms in (\ref{rotHam}) are of the same form as
the spin Hamiltonian (\ref{STRIPEspinEFFECTIVE}) for the ideal
phase-antiphase $(\gamma \!= \!\pi)$ problem, but with a
$\gamma$-dependent coupling $g_{0s}(\gamma)$. Introducing
auxiliary fermion fields $\tilde{R}_{\mu}(x)$ and
$\tilde{L}_{\mu}(x)$, connected to the rotated currents
$j_{L/R}^i(x)$ by
\begin{equation} \label{auxiliary}
j_L^i = \frac{1}{2}\no{\tilde{L}^{\dagger}_{\mu} \sigma^i_{\mu
\nu} \tilde{L}_{\nu}}, \ \ \ \ \ \ \ j_R^i =
\frac{1}{2}\no{\tilde{R}^{\dagger}_{\mu} \sigma^i_{\mu
\nu} \tilde{R}_{\nu}}, \ \ \ \ i=x, y, z \, ,
\end{equation}
an application of the bosonization dictionary in Sec. II.C (with
$L_{\mu} \rightarrow \tilde{L}_{\mu}, R_{\mu} \rightarrow
\tilde{R}_{\mu}, J_{R/L}^z \rightarrow j_{R/L}^z$) produces, as
expected, a sine-Gordon model for the corresponding bosonic spin field
$\tilde{\varphi}_s$ and its dual $\tilde{\vartheta}_s$. 
Using the same procedure to bosonize also the last term in
(\ref{rotHam}) $-$ which is new $-$ we finally obtain
\\
\begin{widetext}
\begin{equation} \label{DSG}
H_s(\gamma) = \frac{v_s}{2} \int dx\Big\{(\partial_{x}
\tilde{\varphi}_{s}^{\prime})^2
+ (\partial_x\tilde{\vartheta}_{s}^{\prime})^2 +
\frac{2m^{(\tilde{\varphi})}_{s}}{a_0^2}\kappa\cos(\sqrt{8\pi
K_{s}}\tilde{\varphi}_{s}^{\prime}) +
\frac{2m^{(\tilde{\vartheta})}_{s}}{a_0^2}\kappa\cos(\sqrt{8\pi
K^{-1}_{s}}\tilde{\vartheta}_{s}^{\prime}) \Big\}, 
\end{equation}
where $-$ to leading order in the coupling constants $-$ we have:
\begin{eqnarray}
2(K_s-1) & = & g_{0s}(\gamma)  = \frac{a_0}{\pi v_F}
\Big(U-2V+
\frac{J_K^2}{2J_H}\mbox{sin}^2(\gamma/2)\Big), \nonumber \\ 
2\pi m^{(\tilde{\varphi})}_s & = & g_{0\perp}(\gamma)   =  \frac{a_0}{\pi v_F}\Big(U-2V+
\frac{J_K^2}{2J_H}(1-\frac{1}{2}\mbox{sin}^2(\gamma/2))\Big), \nonumber \\
2\pi m^{(\tilde{\vartheta})}_s & = & g_{0f}(\gamma)  = 
-\frac{a_0}{\pi v_F}\frac{J_K^2}{4J_H}\mbox{sin}^2(\gamma/2), \nonumber \\
v_s & = & v_F - \frac{a_0}{2\pi}(U-2V+\frac{J_K^2}{J_H}) \,
. \\
& & \nonumber 
\label{DSGparameters}
\end{eqnarray}
\end{widetext}

Thus, the spin sector of the stripe electrons in the presence of twisted
antiferromagnetic domains is described by a {\em generalized sine-Gordon model}
containing a cosine of the dual spin field ${\vartheta}^{\prime}_s$. The
presence of this term allows for spin-nonconserving processes, reflecting 
the complete breaking of the spin rotational symmetry (down to an Ising like 
$Z_2 \times
Z_2 \times Z_2$ discrete symmetry, with one $Z_2$ factor for each spin component
in the rotated basis (\ref{RotatedCurrents})). 
This model was first studied by Giamarchi and Schulz \cite{GS},
who used it to account for spin-orbit and electronic dipole-dipole interactions
in quasi-one-dimensional conductors. Its perturbative RG equations
(valid in the limit of small parameters $g_{0s}(\gamma), g_{0\perp}(\gamma)$
and $g_{0f}(\gamma)$ are most easily derived by exploiting the operator
product expansion \cite{GNT} for the rotated currents in Eq.~(\ref{RotatedCurrents}),
using (\ref{rotHam}), and one obtains:
\begin{eqnarray} \label{DSGRG}
\frac{dg_s}{d\ell} & = & g_f^2 - g_{\perp}^2\, , \nonumber \\ 
\frac{dg_{\perp}}{d\ell} & = & - g_s g_{\perp}\, , \\  
\frac{dg_f}{d\ell} & = & g_s g_f\, . \nonumber
\end{eqnarray}
The renormalized couplings $g_i \equiv g_i(\gamma, \ell), 
(i=s, \perp , f)$ are connected
to the bare parameters $g_{0i}(\gamma)$ in (\ref{DSGparameters}) 
by $g_i(\gamma, \ell = 0) \equiv
g_{0i}(\gamma)$, with $\ell = \mbox{ln} (a/a_0)$, where $a$ 
is a renormalized length.
Focusing on the case with repulsive electron-electron
interaction $(U-2V + J_K^2/2J_H > 0$, with $K_c < 1)$, we read off from
(\ref{DSGparameters}) that $g_{0s}(\gamma) > 0$. For this case
the RG equations (\ref{DSGRG}) support two strong-coupling
massive phases:
\begin{eqnarray}
\label{DSGphase1}
(i) \ g_{0s}(\gamma) < g_{0\perp}(\gamma) \, - \left|
g_{0f}(\gamma)\right| : \nonumber \\ 
g_s \rightarrow -\infty,  \ \left|
g_{\perp}\right| \rightarrow \infty,  \ g_f \rightarrow 0 \, ,
\end{eqnarray}
and
\begin{eqnarray}
\label{DSGphase2}
(ii) \ g_{0s}(\gamma) > g_{0\perp}(\gamma)\, - \left|
g_{0f}(\gamma)\right| : \nonumber \\ 
g_s \rightarrow \infty, \ \left|
g_{f}\right| \rightarrow \infty, \ g_{\perp} \rightarrow 0\, .
\end{eqnarray}
By inspection, using (\ref{DSGparameters}), we can label the two regions
by the range of the twist angle $\gamma$:
\begin{equation}          
\mbox{(i)} \ \ 0 < \gamma < \pi/2\, , \ \ \ \ \mbox{(ii)} \ \ \pi/2 < \gamma < \pi\, .
\end{equation}
In region (i) $\left| g_s \right|$ and $\left| g_{\perp} \right|$ increase
upon renormalization, while $g_f$ scales to zero. The dual spin
field is hence irrelevant and the picture emerging is
qualitatively the same as in the A phase of Sec. III.C.2: the
SDW$^z$ correlation is the leading instability (critical
exponent $K_c$), with TS$^z$ being subleading (critical exponent
$1/K_c$). Note, however, that the ordering tendency is now along the
$z$-direction of the {\em rotated frame}, defined by the transformation
in (\ref{RotatedCurrents}). It follows that for a twist angle
$0 \le \gamma < \pi/2$, the leading
(subleading) instability is that of a spin density wave (triplet
pairing) oriented along a line in the plane spanned by
$\langle \vn^{(A)}\rangle$ and  $\langle \vn^{(B)}\rangle$, and making  
an angle $\gamma/2$ with $\langle \vn^{(A)} \rangle$.

Turning to region (ii), the renormalized coupling constants
$g_s$ and $\left| g_f \right| $ are seen to flow to infinity, while
$g_{\perp}$ scales to zero. For our case, where $g_{0f}(\gamma) 
> 0$, it follows that $\tilde{\vartheta}_s$ gets ordered with
expectation value $\langle \tilde{\vartheta}_s \rangle = 0$, while 
the spin field $\tilde{\varphi}_s$ stays
disordered. By running down the 
list of bosonized order parameters in Sec III.B.1, we
pinpoint the leading (subleading) instability as that of the
SDW$^y$ (TS$^y$) correlation, with critical exponent $K_c \, (1/K_c)$.
Note that also this result refers to the  
{\em rotated frame} defined by (\ref{RotatedCurrents}).
Translating back to the original frame we infer that for a twist
angle $\pi/2 < \gamma \le \pi$, the leading
(subleading) instability is again that of a spin density wave (triplet
pairing) oriented along an axis contained in the
$\langle\vn^{(A)}\rangle - \langle\vn^{(B)}\rangle $-plane, but now making
an angle $\gamma/2+ \pi/2$ with $\langle\vn^{(A)}\rangle $.

Two comments are here appropriate. First note that the axis which defines
the enhanced SDW and TS correlations makes a
$\pi/2$-jump as the twist angle $\gamma$ passes through
$\pi/2$. This may appear reminiscent of a spin-flop transition induced
by a magnetic field, as seen in certain quasi-1D materials 
\cite{GS}.
However, there is no physical effect associated with the
jump other than a sudden change of the direction along which the
SDW and TS correlations are enhanced.
Secondly, when $\gamma = \pi$ we do 
recover the result for the A phase in Sec III.C.2 (ideal phase-antiphase
configuration). Note that the presence of the dual cosine-term in the effective model 
(\ref{DSG}) simply reflects the fact that we are now working in the rotated frame. 
By undoing the rotation in (\ref{RotatedCurrents}) we immediately recover the
standard sine-Gordon model, as given in (\ref{SGs}) for 
a perfect phase-antiphase configuration.

To summarize; we have shown that 
the opening of a $\Neel$ twist angle $\gamma \neq \pi$ 
across the stripe does not remove the instabilities of the A
phase found in Sec. III.C.2. The conventional leading
(subleading) instability is still that of a spin density wave
(triplet pairing), but now tilted w.r.t. to the two $\Neel$-
directions. 

By inspection of the {\em composite} order parameters defined in 
(\ref{COMPCDW})
and (\ref{CSP}) it is clear that their amplitudes will drop as
$\gamma$ decreases from its largest value $\gamma = \pi$, defining a
perfect phase-antiphase $\Neel$ configuration. It is here
interesting to note that when $\gamma \neq \pi$ there is room for
additional composite order parameters. Besides the trivial
variations of (\ref{COMPCDW}) and (\ref{CSP}) where $\vS^{(A)} -
\vS^{(B)} \rightarrow \vS^{(A)} + \vS^{(B)}$, we can construct
two {\em composite vector order parameters:}
\begin{eqnarray} 
{\vO}_{{\small c-SDW}} & \sim & \vO_{SDW} \times ( \vS^{(A)} \times
\vS^{(B)} ) \\ \nonumber
  & \rightarrow & \vO_{SDW} \times ( \vn^{(A)} \times
\vn^{(B)} ) \label{compSDW}\, ,
\end{eqnarray}
and
\begin{eqnarray} 
{\vO}_{{\small c-TS}} & \sim & \vO_{TS} \times ( \vS^{(A)} \times
\vS^{(B)} ) \\ \nonumber
   & \rightarrow & \vO_{TS} \times ( \vn^{(A)} \times
\vn^{(B)} ) \, , \label{compTS}
\end{eqnarray}
where in the second lines we have dropped the small contributions
from the $\vl$-fields (which can be neglected in the correlation functions,
cf. Sec. III.B).
These composites have the same symmetries as the corresponding conventional 
order parameters, SDW and TS, respectively.
In particular, ${\vO}_{{\small c-TS}}$ is even under time and spin reversal, 
while odd under a parity transformation. It follows that for a sufficiently
rigid spin texture with $\gamma$ away from $0$ and $\pi$, triplet pairing comes
in two guises: one conventional mode and one composite
mode, both carrying zero momentum. These pairing modes have
enhanced correlations along well-defined directions, orthogonal
to each other. 
As $\gamma \rightarrow \pi$ (perfect antiphase boundary)
or $\gamma \rightarrow 0$ (''in-phase'' boundary), the composite 
triplet channel gets suppressed 
and one is left with the pairing instabilities characterizing the A phase in Sec. III.
Elsewhere we shall explore the consequences of this intriguing 
possibility.

\section{Summary}

In this article we have analyzed the problem of a one-dimensional
electron liquid {\em (charge stripe)} embedded in a
two-dimensional antiferromagnetic insulator, and coupled to it
via a weak spin exchange. Using a mean-field type construction,
the spin exchange gets encoded as an effective anisotropic
spin interaction among the electrons. This interaction is shown to
be marginally relevant under certain conditions (in a renormalization
group sense), producing enhanced pairing and
density fluctuations in the electron liquid.

For realistic values of the model parameters - assuming a
screened Coulomb interaction for the electrons, and a
conduction band away from half-filling - the dominant 
instabilities are towards a
conventional spin density {\em and} a composite charge density
wave, coexisting with subleading conventional triplet {\em and}
composite singlet pairing correlations. Taking into account the
slow transverse fluctuations of a stripe, the triplet pairing
instability is expected to turn into the dominant one.  While the
magnitudes of the conventional instabilities do not change with
the relative orientation of the $\Neel$ directions in the two
domains surrounding the stripe, the composite correlations have
largest amplitudes when the stripe forms an antiphase domain wall
in the antiferromagnet. With the possible exception of the
composite singlet pairing mode, the instabilities are found to be
insensitive to the spatial orientation of the stripe on the
underlying lattice {\em (collinear} or {\em diagonal)}.   

Our study has been motivated by a wish to understand the role of
spin exchange between stripes and their environment in the
cuprate superconductors. As is well-known, these materials are
exceedingly complex systems. It is unlikely that a simplified
model like ours - where several aspects of the problem have been
simply ignored - can produce accurate predictions about
experimental results. We have discussed {\em a posteriori} some
of the effects not included in our model: the role of slow stripe
fluctuations and interstripe couplings, the finite width of the
antiferromagnetic domains encountered in real materials, as well
as the expected poor screening of the Coulomb interaction along a stripe.
Other aspects of the problem are yet to be addressed; most
importantly the possible appearance of a retarded nonlocal spin
interaction coming from large-amplitude fluctuations of the
$\Neel$ order parameters. Also, the finite-size and boundary
effects implied by the mesoscopic scale at which the stripes
live need to be carefully studied.

The virtue of our ''stripped-down'' model is that it allows us to
carry out a {\em
well-controlled analytical study}, and as such it could serve as 
a stepping stone for more detailed investigations.
The model predicts 
that a stripe-environment spin exchange under certain conditions may
produce instabilities towards pairing of electrons. This result is
obtained {\em without the assumption of a pre-existing spin gap in the
insulating environment}. When including effects from the fluctuations
of the stripes the 
symmetry of the dominant pairing instability appears to come out
''wrong'' $-$ as judged against available experimental
evidence from the cuprates. Does this imply that a stripe-environment spin exchange
plays no role for superconductivity in these materials? Or, could
the picture change when properly building more dynamical elements
into the model? To answer these questions will require more work. 

\acknowledgments

We thank S.~\"Ostlund for a contribution on spin textures in two-dimensional
antiferromagnets, and for kindly allowing us to use it in Sec. IV. 
It is also a pleasure to thank I.~Affleck, N.~Andrei, M.~Granath, 
C.~Morais Smith, B.~Normand, and J.~Voit for helpful discussions.
GIJ acknowledges support by the SCOPES grant N 7GEPJ62379.
The work of HJ was supported by the Swedish Research Council, grant 
621-2002-4947.
\newpage



\begin{thebibliography}{99}
\bibitem{OddFrequencyGap} Note, however, that an odd-frequency gap function itself cannot serve as an 
order parameter since its equal-time expectation value vanishes. Rather, some moment
of the gap function with frequency could be an appropriate order parameter.
\bibitem{chargegap} Note that the viability of pair hopping in addition requires that the {\em charge gap}
of the insulating environment is sufficiently small so as not to obstruct the process.
\bibitem{FOOTNOTE2} The property that only the $\vl$-field couples manifestly to the stripe 
also for a half-filled band uniquely depends on the stripe being "topological" 
(i.e. separating two phase-antiphase N\'{e}el domains) and does not apply to the standard
Heisenberg-Kondo lattice model 
\cite{TsvelikReview} or its extension in Ref. 16.
\bibitem{SDWUmklapp} Note that by coupling $ \tilde{{\cal
O}}^{\dagger}_{{SDW}^z}(j, x)$ and  $\tilde{{\cal O}}_{SDW}^z(j +
1, x)$ we eliminate Umklapp processes that would have
resulted from the coupling of the ''full'' SDW$^z$ order
parameters ${\cal O}^{\dagger}_{{SDW}^z}(j, x)$ and  ${\cal
O}_{SDW^z}(j + 1, x)$; cf. Eq.~(\ref{bosSDWz}).
\end{thebibliography}
\end{document}